\documentclass[traditabstract]{aa} 
\usepackage{graphicx}
\usepackage{txfonts}
\usepackage{rotating}
\usepackage{multirow}
\usepackage{url}
\usepackage{txfonts}
\usepackage{pstricks}
\usepackage{natbib}
\usepackage{subfigure}
\usepackage{enumerate}
\usepackage{slashbox}

\newcommand{\avgg}[1]{\left< #1 \right>} 
\newcommand{\micron}{\mu {\rm{m}}}

\begin{document}

 \title{Mid-infrared microlensing of accretion disc and dusty torus in quasars: effects on flux ratio anomalies\thanks{Microlensing maps and source profiles used for our simulations are available in electronic form at the CDS via anonymous ftp to \texttt{cdsarc.u-strasbg.fr} (130.79.128.5) or via {\texttt{http://cdsweb.u-strasbg.fr/cgi-bin/qcat?J/A+A/}} }}

 \author{D. Sluse
 \inst{1} 
 \and M. Kishimoto
 \inst{2}
 \and T. Anguita
 \inst{3,4,5}
 \and O. Wucknitz 
 \inst{1,2}
 \and J. Wambsganss
 \inst{6}
 }

 \institute{{Argelander-Institut f\"ur Astronomie, Auf dem H\"ugel 71, 53121 Bonn, Germany.}
 \email{dsluse@astro.uni-bonn.de}
 \and 
 {Max Planck Institut f\"ur Radioastronomie, Auf dem H\"ugel 69, 53121 Bonn, Germany.}
 \and
 {Centro de Astro-Ingenier\'ia, Departamento de Astronom\'ia y Astrof\'isica, Pontificia Universidad Cat\'olica de Chile, Casilla 306, Santiago, Chile.}
 \and 
 {Max-Planck-Institut f\"ur Astronomie, K\"onigstuhl 17, 69117 Heidelberg, Germany.}
 \and
 {Departamento de Ciencias Fisicas, Universidad Andres Bello, Av. Republica 252, Santiago, Chile.}
 \and
 Astronomisches Rechen-Institut am Zentrum f\"ur Astronomie der Universit\"at Heidelberg M\"onchhofstrasse 12-14, 69120, Heidelberg, Germany.
  }

 \date{Received: December 4, 2012; Accepted: March 1, 2013 }

  \abstract
 {Multiply-imaged quasars and active galactic nuclei (AGNs) observed in the mid-infrared (MIR) range are commonly assumed to be unaffected by the microlensing produced by the stars in their lensing galaxy. In this paper, we investigate the validity domain of this assumption. Indeed, that premise disregards microlensing of the accretion disc in the MIR range, and does not account for recent progress in our knowledge of the dusty torus which have unveiled relatively compact dust emission. To simulate microlensing, we first built a simplified image of the quasar composed of (i) an accretion disc whose size is based on accretion disc theory, and (ii) a larger ring-like torus whose radius is guided by interferometric measurements in nearby AGNs. The mock quasars are created in the $10^{44.2}-10^{46}$ erg/s (unlensed) luminosity range, which is typical of known lensed quasars, and are then microlensed using an inverse ray-shooting code. We simulated the wavelength dependence of microlensing for different lensed image types and for various fractions of compact objects in the lens. This allows us to derive magnification probabilities as a function of wavelength, as well as to calculate the microlensing-induced deformation of the spectral energy distribution (SED) of the lensed images. We find that microlensing variations as large as 0.1\,mag are very common at 11\,$\micron$ (typically rest-frame 4\,$\micron$). The main signal comes from microlensing of the accretion disc, which may be significant even when the fraction of flux from the disc is as small as 5\% of the total flux. We also show that the torus of sources with $L_{\rm {bol}} \lesssim 10^{45}$\,erg/s is expected to be noticeably microlensed. Microlensing may thus be used to get insight into the rest near-infrared inner structure of AGNs. Finally, we investigate whether microlensing in the mid-infrared can alter the so-called $R_{\rm cusp}$ relation that links the fluxes of the lensed images triplet produced when the source lies close to a cusp macro-caustic. This relation is commonly used to identify massive (dark-matter) substructures in lensing galaxies. We find that significant deviations from $R_{\rm cusp}$ may be expected, which means that microlensing can explain part of the flux ratio problem. }

 \titlerunning{Mid-infrared microlensing of the dusty torus and accretion disc in quasars}
 \authorrunning{D. Sluse et al.}

  \maketitle

\section{Introduction}
\label{sec:intro}

Gravitational lensing of quasars is now a well established tool used for a wide variety of astrophysical and cosmological applications \citep{Claeskens2002, koch2004, Treu2010, Bartelmann2010}. Mass models of lensing galaxies generally reproduce the lensed image configuration down to milli-arcsec accuracy \citep{Chantry2010, Sluse2012a}, but often fail to reproduce the flux ratios between the lensed images \citep{Yonehara2008}. Indeed, because the light rays from the different lensed images ``cross'' different regions of the lensing galaxy, they can get reddened by a different amount in each image \citep{Jean1998, Falco1999}. They can also be subject to the microlensing produced by the stars in the lensing galaxy \citep{Chang1979, Irwin1989}. On the other hand, the intrinsic variability of the source, combined with the time delay between the lensed images, introduces spurious changes in the flux ratios when a system is observed at a single epoch. A flux ratio anomaly is assessed when none of these scenarios account for the observed flux ratios. 

A popular explanation for the origin of the anomaly is extra (de)-magnification of one lensed image due to the presence of a clump of mass typically $10^4-10^8\,M_{\sun}$ located close in projection to that image. This effect, often called milli-lensing or meso-lensing \citep{Wambsganss1992, Baryshev1997, Mao1998}, is nowadays an important tool for probing the amount of massive substructures in galaxies \citep{Metcalf2001, Dalal2002, Evans2003, Keeton2003, Vegetti2010, Metcalf2011, Xu2011, Inoue2012}. To use this technique, it is necessary to exclude the other possible causes of flux perturbations. The main one is microlensing. It is fundamentally of the same nature as milli-lensing, but it varies on time scales of months or years (instead of at least centuries), and only affects relatively compact sources of projected size similar\footnote{Detectable microlensing can be observed in sources with projected sizes 10 times larger than $\eta_0$ \citep{Refsdal1991, Refsdal1997}.} to or smaller than one microlens Einstein radius $\eta_0 \sim 0.01\,$pc (Sect.~\ref{sec:micro}). For this reason, flux ratio anomalies detected at radio wavelengths are considered as the most robust because the sources are believed to be larger than $\eta_0$ \citep[but see][]{Koopmans2000}. Unfortunately, radio flux ratios are measurable only for a small sub-sample of lensed quasars bright enough at radio wavelengths (about one fifth of the known objects), and other proxies of the intrinsic flux ratios are desirable. 

Flux-ratios measured at 8--11\,$\micron$ (corresponding to $\sim$\,2.9--4\,$\micron$ rest-frame for typical lensed AGN), are considered as a promising alternative to estimate unbiased flux ratios because most of the emission is assumed to originate from the large size torus \citep{Chiba2005, Minezaki2009}. Recent progress in our knowledge of the dust tori of quasars and active galactic nuclei question the generality of this assertion. 

The distance of the dust torus to the central engine of nearby AGNs has been obtained using reverberation-mapping, which infers the distance to the central black hole from the measurement of the time-lag between optical and near-infrared (NIR) flux variations \citep[][and references therein]{Glass1992, Suganama2006}. \cite{Suganama2006} compiled the available inner radii of the dust tori at 2.2\,$\micron$ for 10 objects, and have found that the reverberation mapping radii $R_{\tau_K}$ is consistent with an increase with $L^{0.5}$, in agreement with the expected dependence of the dust sublimation radius with luminosity \citep{Barvainis1987}. The reverberation-radius is supposed to be a good approximation of the dust sublimation radius, but could be larger depending on the exact responsiveness of the dust to the incident continuum flux. However, the measured reverberation radius is actually smaller by a factor of $\sim$ 3 \citep{Kishimoto2007} than the naive expectation for the sublimation radius using dust properties representative for ISM dust grains \citep{Barvainis1987}. Near-infrared (NIR) interferometry has also revealed that the NIR emitting region is relatively compact, and that the radius of the torus can be as small as the typical Einstein radius of a microlens \citep{Swain2003, Kishimoto2011a}. In addition, we know that in the NIR range, there is still a small fraction of the emission originating from the accretion disc of the AGN. This means that microlensing could be important in the NIR-MIR range. 

\begin{figure}
\centering
\begin{tabular}{c}
\includegraphics[scale=0.5]{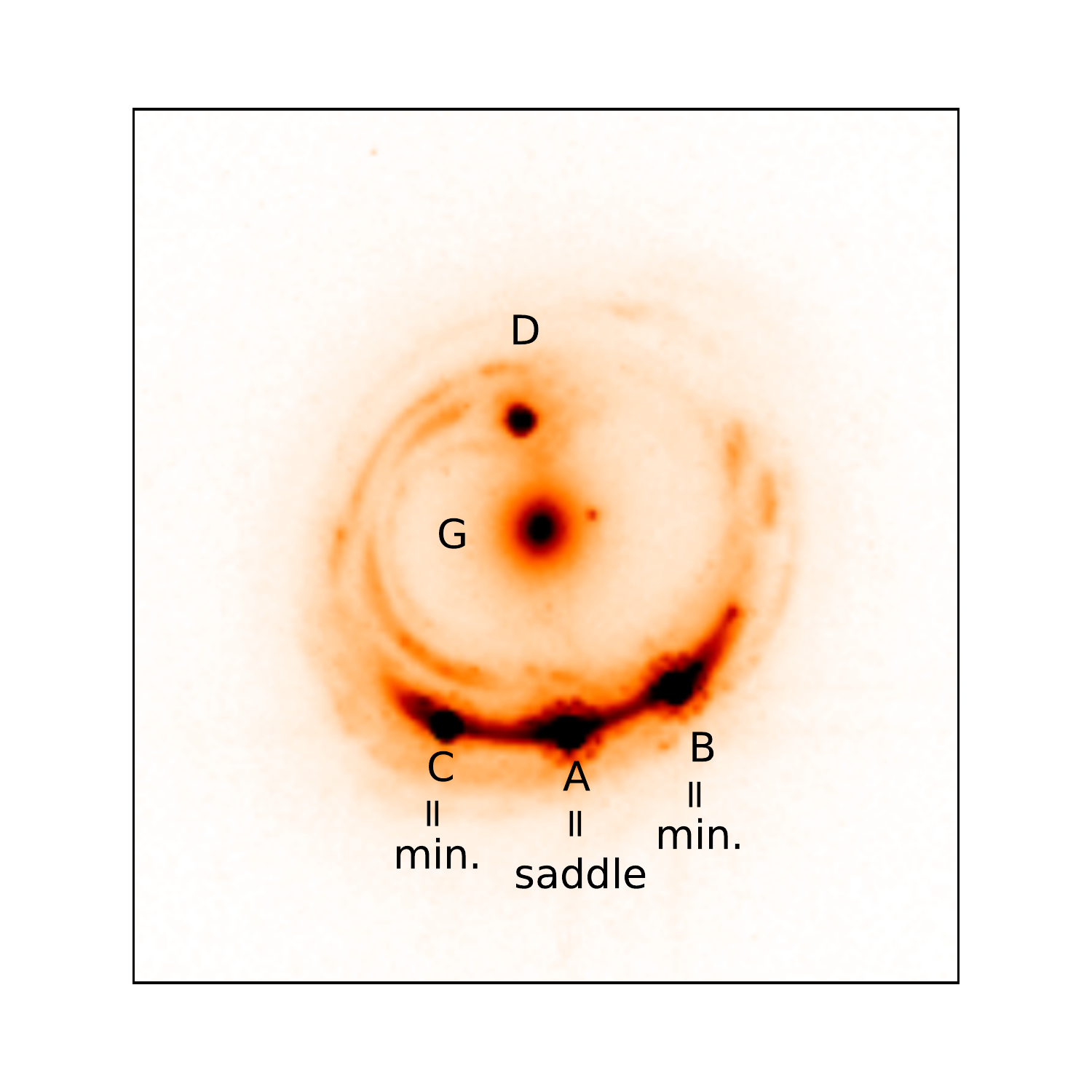}\\
\end{tabular}
\caption{Example of quadruply-imaged lensed quasar where the lensed images are in ``cusp configuration'' \citep[J1131$-$1231, from][]{Claeskens2006}. In such a system, the flux of image $A$ (saddle-point of the arrival time surface) is expected to be equal to the sum of the fluxes of $B$ and $C$ (minima of the arrival time surface).}
\label{fig:cusp}
\end{figure}

Different mass models of the lensing galaxy which reproduce the lensed images position, often lead to different predicted flux ratios. It is therefore not sufficient to have a flux measurement free of extrinsic contamination to identify a flux ratio anomaly. Fortunately, for some particular lens-source configurations, the flux ratio between lensed images can be predicted in a nearly model-independent way. In particular, for the so called ``cusp-configuration'' systems (Fig.~\ref{fig:cusp}), the unsigned magnification of the saddle-point lensed image{\footnote{Lensed images are created at a minima, a maxima or saddle point(s) of the arrival time-surface associated the wavefront originating from the source. The image parity is calculated based on the magnification tensor \citep[e.g.][]{Blandford1986, Schneider1992}.}} is expected to be equal to the sum of the magnifications of the two (adjacent) minimum images. After the seminal work of \cite{Mao1998}, it has been realised that this relation breaks down in the presence of massive substructures in the line-of-sight towards one of the lensed images, providing a simple technique to identify flux ratio anomalies \citep[e.g.][]{Bradac2002, Keeton2003}. 

In this paper we investigate the effect of microlensing on the NIR-MIR emission of lensed quasars accounting for the refined picture we have for the dust torus. We also explore the role of the accretion disc emission contributing a measurable fraction of the MIR-NIR emission. In Sect.~\ref{sec:source}, we present our model of the source. In Sect.~\ref{sec:micro}, we explain how we proceed to perform the microlensing simulations. The results of our simulations are given in Sect.~\ref{sec:results} and discussed in  Sect.~\ref{sec:discussion}. In Sect.~\ref{sec:fluxratios}, we investigate if the amount of microlensing taking place at MIR wavelengths can affect the detection of flux ratio anomalies produced by substructures, with a particular emphasis on their influence on the so called $R_{\rm {cusp}}$ relation. In Sect.~\ref{sec:SED}, we study the effect of microlensing on the SED of the lensed images. Finally we conclude in Sect.~\ref{sec:conclusions}. 

\section{Brightness profile of the source}
\label{sec:source}

The continuum emission from AGNs in the range 1--5.5 $\micron$ (rest-frame) comes from two components: (i) the emission originating from the compact accretion disc, and (ii) the extended emission associated with the dust torus. In Sect.~\ref{subsec:cont} \&~\ref{subsec:dust}, we introduce the relation used to model the brightness profiles of these two components. In Sect.~\ref{subsec:composite}, we explain  how we fix the relative contribution of each component and how this varies with wavelength. 

\subsection{Accretion disc emission}
\label{subsec:cont}

We consider that the accretion follows a standard accretion disc model with $f_{\nu} \propto \nu^{1/3}$ \citep{Shakura1973}. \cite{Kishimoto2008} showed observational evidence of the validity of this model in the near-infrared range. We assume the validity of this model up to $\sim$ 5 $\micron$ (rest-frame). Within this model, the half-light radius{\footnote{We assume $R^{AD}_{1/2} = 2.44\,R_{\lambda}$, with $R_{\lambda}$ the disc size. This relation is only an approximate relation strictly valid if the disc inner radius $R^{\rm {AD}}_{\rm{in}} << R^{AD}_{1/2}$ \citep[See e.g.][ Sect. 2.3]{Abolmasov2012}}} of the disc is given by: 

\begin{equation}
R^{AD}_{1/2} (\lambda) = 2.37\,10^{16}\,\rm{cm}\,\left(\frac{\lambda_{\rm {rest}}}{\micron}\right)^{4/3} \, \left(\frac{M_{\rm {BH}}}{10^9 M_{\sun}}\right)^{2/3} \, \left(\frac{L_{\rm {bol}}}{\eta_{\rm {eff}}\,L_{\rm {edd}}} \right)^{1/3} 
\label{equ:cont}
\end{equation}

\noindent where $\lambda_{\rm {rest}}$ is the rest-wavelength of the emission, $M_{\rm {BH}}$ is the mass of the central black hole, $L_{\rm {bol}}$ is the bolometric luminosity of the object and $\eta_{\rm {eff}}$ is the accretion efficiency of the disc. In the following, we assume a typical accretion rate efficiency $\eta_{\rm {eff}}=0.1$, and a typical Eddington ratio $L_{\rm {bol}}/L_{\rm {edd}} = 1/3$. Since we fix the Eddington ratio, we can derive $M_{\rm {BH}}$ from $L_{\rm {edd}}$. Under these assumptions, the size of the continuum (accretion disc) emission is proportional to the bolometric luminosity $L_{\rm {bol}}^{2/3}$.

Some microlensing studies of the X-ray to optical size of the accretion disc suggest that: (i) the optical accretion disc is larger by one order of magnitude than the accretion disc prediction, and (ii) the size of the disc $R^{AD}_{1/2} \propto \lambda^{1/\beta}$, with $\beta > 3/4$, in disagreement with the accretion disc theory \citep{Pooley2007a, Morgan2010a, Blackburne2011a}. However, other microlensing studies agree well with accretion disc theory \citep[e.g.][]{Bate2008, Eigenbrod2008b, Floyd2009a}. Possible explanations to these results involve deviations of some discs from the standard disc, under-estimated Eddington ratios, ``contamination'' of the optical continuum emission by reprocessed and/or scattered light from the disc, systematic errors associated to the analyses \citep{Abolmasov2012, Lawrence2012}. However, these results concern the optical range, and to date, there has been no claim of discrepancy in the (less well explored) NIR-MIR range \citep{Agol2009}.

\subsection{Dust torus}
\label{subsec:dust}

The dust torus of the quasar is modelled as a ring-like structure. For the inner radius of the ring, we adopt the reverberation radius in rest-frame $K$-band (2.2\,$\micron$) measured by \cite{Suganama2006}. Following Equ.~3 of \cite{Kishimoto2007}, which quantifies the fit of the reverberation radius as a function of $L$ done by \cite{Suganama2006}, we can relate the inner radius of the torus to the UV luminosity of the AGN: 

\begin{equation}
R_{\rm{in}} = 0.47\,\left(\frac{6\,\nu\,L_{\nu}(5500\AA)}{10^{46}\,{\rm{erg/s}}}\right)^{1/2} {\rm {pc}}.
\label{equ:RtauK}
\end{equation}

\noindent We use $L_{\rm {bol}}\,=\,$9.26\,$\nu\,L_{\nu}(5500 \AA) $ \citep{Shen2008} to derive the bolometric luminosity. 

The other two parameters that define our model are the surface brightness profile and the outer radius $R_{\rm {out}}$ of the torus. The surface brightness profile of the torus is poorly known. Theoretical considerations support a power-law surface brightness \citep{Barvainis1987} but current data do not allow one to properly test this prediction over the NIR (rest) regime. This is however not very important for our purposes, because microlensing depends only weakly on the exact surface brightness as far as the same half-light radius of brightness profile is used \citep{Mortonson2005}. We therefore use a uniform profile between $R_{\rm {in}}$ and $R_{\rm {out}}$. Most of the simulations are done for a face-on disc. We comment in Sect.~\ref{subsec:PA} on the effect of inclination. Interferometric measurements of local Type 1 AGNs show that the torus is relatively compact at 2.2 $\micron$, with $R_{\rm{out}}/{R_{\rm{in}}} \leq 2$ \citep[Fig.\,7 of][]{Kishimoto2011b}. At larger wavelengths (i.e. 8.5 and 13 $\micron)$, the torus tends to be less compact (except for the brightest objects) and $R_{\rm{out}}/R_{\rm{in}} > 5$ is conceivable. Since our observational knowledge on the change of the size of the torus both with wavelength and luminosity is still quite limited, we choose not to parametrise the variation of the size of the torus with wavelength. Instead, in this work we consider two extreme scenarios. A compact torus with $R_{\rm{out}}/{R_{\rm{in}}} = 1.2$ and a more extended torus with $R_{\rm{out}}/{R_{\rm{in}}} = 5$. Under these assumptions, half of the light of the ring is between $R_{\rm{in}}$ and $R^{{\rm{torus}}}_{1/2} \sim 1.1\,R_{\rm{in}}$ for the compact torus model and $R^{{\rm{torus}}}_{1/2} \sim 3.6\,R_{\rm{in}}$ for the extended model.

\subsection{Compound model}
\label{subsec:composite}

\begin{figure}

\centering
\includegraphics[scale=0.48]{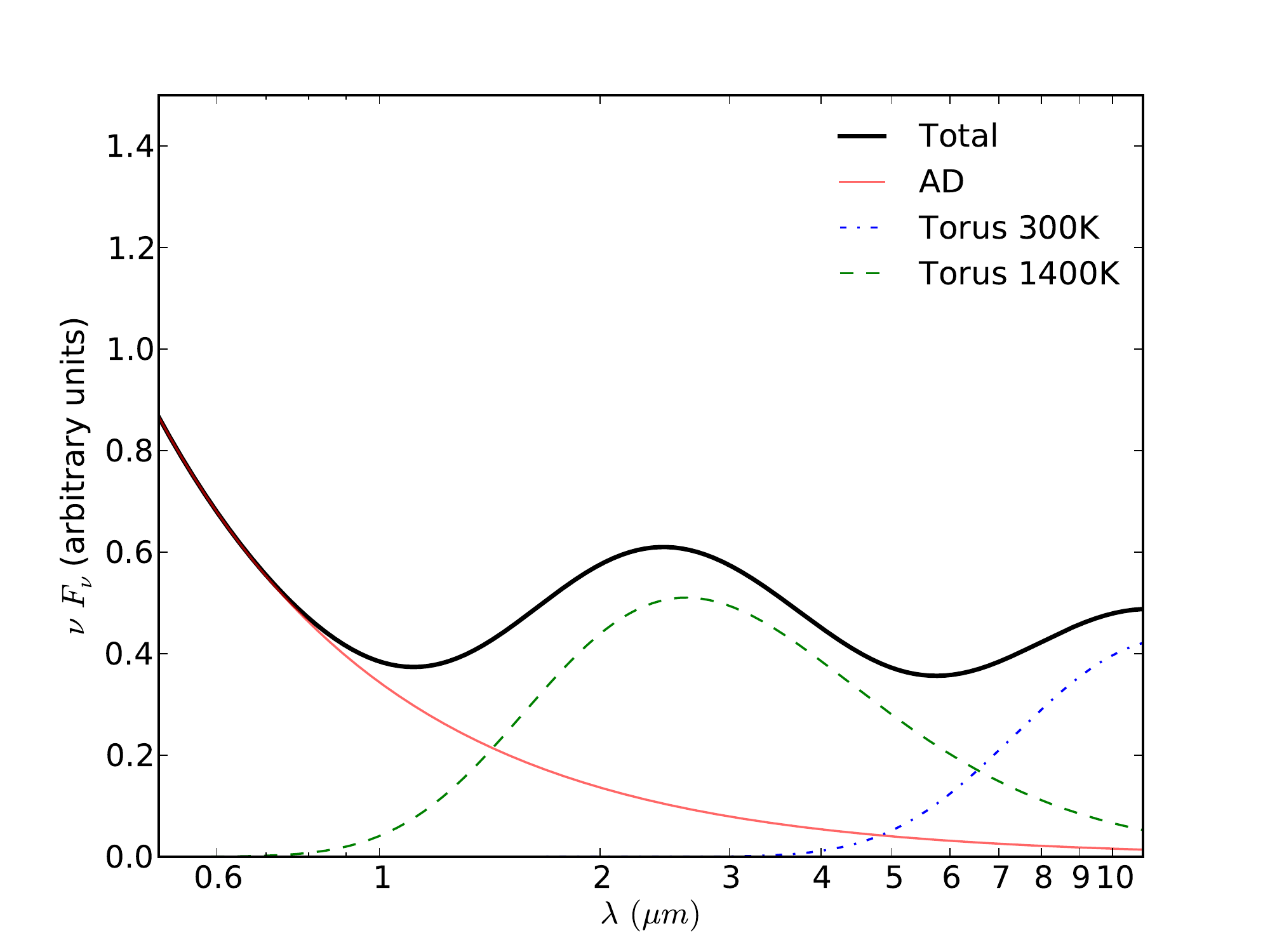}
\caption{Spectral Energy Distribution (SED) of our source model between 0.5 and 11\,$\micron$ (rest-frame). The accretion disc (AD) emission is shown with a solid red line, the ''hot'' component of the torus with a dashed green line, the ''cold'' component with dashed-dotted blue line and the total SED with thick black line.}
\label{fig:SED}
\end{figure}

In order to build a realistic model of the quasar source, it is necessary to estimate what fraction of the accretion disc contributes to the total flux at a given wavelength. The modelling of the SED of the nearby AGNs used for interferometric studies suggests that the continuum contributes 14--29\% of the total emission at 2.2\,$\micron$ \citep{Kishimoto2011a}. Therefore, we assume a fiducial contribution of the accretion disc to the flux of $f_{AD}=$\,0.2 at 2.2\,$\micron$. In order to derive $f_{AD}$ at other wavelengths, we use $f_{\nu} \propto \nu^{1/3} $ for the accretion disc and we consider a sum of two black-body emissions for the dust. The use of two black-body components (a ''high'' and a ''low'' temperature component) for the dust emission provides a reasonably good description of the near-infrared SED of Type 1 AGNs. We use a ''high'' temperature component BB1 with a fiducial temperature $T=1400$\,K, and a ``low'' temperature component BB2 with $T=300$\,K, which dominates the dust emission at $\lambda > 7\,\micron$. The relative contribution of each component is given by the ratio between the effective area of the corresponding emitting region. Interferometric measurements by \cite{Kishimoto2011a} indicate that the effective area of the BB2 region emission is 100--1000 times larger than the BB1 region emission \citep[Table 7 of ][]{Kishimoto2011a}. We choose a typical ratio of effective area $\Omega_2/\Omega_1 = 400$ leading to the following parametrisation of the torus flux $F_{\nu}^{\rm {torus}} = 400\,B_{\nu}(\nu, T=300\,{\rm {K}})+B_{\nu}(\nu, T=1400\,{\rm {K}})$. Our model of the spectral energy distribution of the source is shown in Fig.~\ref{fig:SED}.

In the following, we investigate the effect of microlensing at rest-frame wavelengths of 1, 2.2, and 4.4\,$\micron$. We choose these wavelengths because they roughly correspond to the observed $K-$band, $L-$band and 11\,$\micron$ of the lower redshift lensed quasars (1.1 $\lesssim z \lesssim $ 1.5). At these wavelengths, the contribution of the cold $T=300$\,K component of the dust remains less than a few percent. It is therefore unnecessary to model explicitly its surface brightness separately. We only include it in the calculation of $f_{AD}(\lambda)$, to account for the fact that the accretion disc becomes more and more diluted with increasing wavelength. 

Most of the known lensed quasars have unlensed bolometric luminosities $L_{\rm {bol}}$ in the range [$10^{44}-10^{46.5}$] erg/s. Therefore, we decided to study microlensing for 3 different luminosities: $L_{\rm {low}} = 10^{44.2}$\,erg/s, $L_{\rm {med}} = 10^{45}$\,erg/s and $L_{\rm {high}} = 10^{46}$\,erg/s. We show in Table~\ref{tab:size} the characteristic size of the source, in units of the Einstein radius $\eta_0$ (Eq.~\ref{equ:RE}), for these luminosities. 

\begin{table}[tb]
\caption{Sizes of the accretion disc (AD) and of the dust torus used in our model for three different luminosities (col.\,\#1). Col.~\#2 displays the half-light radius of the accretion disc (in units of the Einstein radius $\eta_0$ for $\avgg{M}=0.3\,M_{\sun}$). The next columns characterise the properties of the torus. Col.~\#3, 4, 5 give the type of torus, its inner radius $R_{\rm{in}}$, and the width of the ring. These quantities are given for $\lambda_{\rm {rest}} = 2.2\,\micron$. Only $R^{\rm {AD}}_{1/2} \propto\lambda^{4/3}$ varies with the wavelength (Eq.~\ref{equ:cont}).} 
\begin{center}
\begin{tabular}{cccccc}
\hline 
$L_{\rm{bol}}$ (erg/s) & $R^{\rm {AD}}_{1/2}$ ($\eta_0$) & torus type & $R_{\rm{in}}$ ($\eta_0$) & $R_{\rm{out}}-R_{\rm{in}}$ ($\eta_0$) \\
\hline
$10^{46}$	& 0.3	& Compact   &	   38	 &   7.6	\\  
                & 	& Extended  &	   38	 & 155.0	\\
$10^{45}$	& 0.06	& Compact   &	   12	 &   2.4	\\
	        & 	& Extended  &	   12	 &  48.0	\\
$10^{44.2}$	& 0.02	& Compact   &	    5	 &   1.0   \\
	        & 	& Extended  &	    5	 &  20.0   \\

\hline 

\hline
\end{tabular} 
\label{tab:size}
\end{center} 
\end{table}

\section{Microlensing}
\label{sec:micro}

The scale-length at which the source is typically affected by microlensing is the Einstein radius of the microlenses, given by: 

\begin{equation}
\eta_0 = \sqrt{\frac{4G\avgg{M}}{c^2}\frac{D_{\rm {os}}D_{\rm {ls}}}{D_{\rm {ol}}}} \sim 3\times10^{16}\,{\rm {cm}}\,\sqrt{\frac{\avgg{M}}{0.3\,M_{\sun}}},
\label{equ:RE}
\end{equation}

\noindent where the $D$ are angular diameter distances, and the indices o, l, s refer to observer, lens and source, respectively, and ${\avgg{M}}$ is the average mass of the microlenses. The value $\eta_0 \sim $ 3 $\times 10^{16}\,$cm\,$\sqrt{{\avgg{M}}/{0.3\,M_{\sun}}} \sim$ 0.01\,pc is not derived for precise source and lens redshifts but is roughly the average $\eta_0$ obtained for known lensed quasars \citep{Mosquera2011}. All along this work, we use this typical value to convert linear source sizes (derived in cm) into Einstein radii.  

Microlensing of the source is estimated in the following way{\footnote{The microlensing maps and source profiles used in this work are made available via CDS.}}. First we build a micro-magnification map using the inverse-ray-shooting code developed by \cite{Wambsganss1990, Wambsganss1999, Wambsganss2001}. This map gives, for a pixel-sized source, the micro-magnification as a function of the source position. Since we want to study microlensing affecting a source with both a compact (continuum) and an extended (torus) structure, we need maps with large extent and relatively high resolution. In practice, we use maps of 100\,$\eta_0$$\times$100\,$\eta_0$, with a spatial resolution of 0.01\,$\eta_0$/pixel for $L_{\rm{bol}}< 10^{45}$\,erg/s and maps of 250\,$\eta_0$$\times$250\,$\eta_0$, with a spatial resolution of 0.05\,$\eta_0$/pixel otherwise. Second, following the methodology described in Sect.~\ref{sec:source}, we build the surface brightness of the source on a grid as fine as the microlensing map. Third, we convolve our image of the source with the microlensing map. Each pixel of that convolved map provides the amount of micro-magnification associated to the extended sources as a function of position. The spatial resolution we considered fixes the lower bound of our luminosity range, $L_{\rm {bol}} = 10^{44.2}$\,erg/s. For that source, we have $R_{1/2}=0.02\,\eta_0$ at 2.2\,$\micron$. Our lower resolution microlensing maps are needed to be able to estimate the amount of microlensing of the extended torus. These maps are however still too small to calculate microlensing for the extended torus when $L = 10^{46}$\,erg/s. In that case, $R_{\rm{out}}-R_{\rm{in}} > 100\,\eta_0$. This size of the torus is too large to be affected by significant microlensing and we therefore assume no microlensing of that model of the torus. 

The amplitude and the probability of (de)-magnification of a source because of microlensing depends on the image type, or more precisely, on the convergence $\kappa$ and shear $\gamma$ at the location of the lensed images. Microlensing also depends on the fraction of matter in smooth and compact form at the projected location of the lensed images. Since we cannot sensibly explore the whole range of plausible values of $\kappa$ and $\gamma$, we have decided to focus on values of $\kappa$ and $\gamma$ typical of saddle-point and minimum image observed in cusp-like systems. Our fiducial saddle-point image has $(\kappa, \gamma) =$ (0.47, 0.57), and our fiducial minimum has $(\kappa, \gamma)=(0.42, 0.50$), corresponding to an average macro-magnification $|\mu_{sad}| = 19.8 $ and $|\mu_{min}| = 11.8$. These values match those obtained for image A and B of the lens J1131$-$1231 (Fig.~\ref{fig:cusp}) using an elliptical Singular Isothermal Ellipsoid + external shear (SIE+$\gamma$) to model the lensing galaxy \citep{Sluse2012a}. Recent microlensing studies of X-ray flux ratios in lensed quasars indicate that the most likely fraction of compact matter in lensing galaxies is $\kappa_* = \kappa_c/\kappa=0.07$, at the location of the lensed images \citep{Pooley2012}. This is the fiducial value we use in this paper. Because there is evidence for a larger fraction of compact objects in some systems, we also test for $\kappa_*=0.3$ and $\kappa_*=1$. 

\section{Results}
\label{sec:results}

The amount of microlensing affecting a macro-lensed image of flux $F_\lambda$ is quantified by the factor $\mu$, which accounts for the extra-magnification induced by microlensing such that $F^{\rm{obs}}_\lambda = \mu\,F_\lambda$. Hereafter, following e.g. \cite{Wambsganss1992a}, we calculate for various source and lens properties, the probability density function ({\it {PDF}}) for the quantity $\Delta m = 2.5\,\log(\mu)$. With that convention, $\Delta m > 0$ ($\mu > 1$) corresponds to a {\it{magnification}} of the source and $\Delta m < 0$ is a {\it {demagnification}}.

For a fixed lensed image property (corresponding to a given combination of $\kappa,\gamma$), three sets of parameters influence the probability of microlensing in the NIR-MIR range: (a) the physical characteristics of the source, namely its luminosity, and the compactness of the dust torus; (b) the relative fraction of compact and smooth dark matter in front of the lensed image; (c) the variable contribution with wavelength of flux originating from the torus and from the disc.

\subsection{Changing the source properties}
\label{subsec:source}

\begin{figure*}
\centering
\setcounter{subfigure}{0}
\subfigure[Saddle-point image]{\includegraphics[scale=0.45]{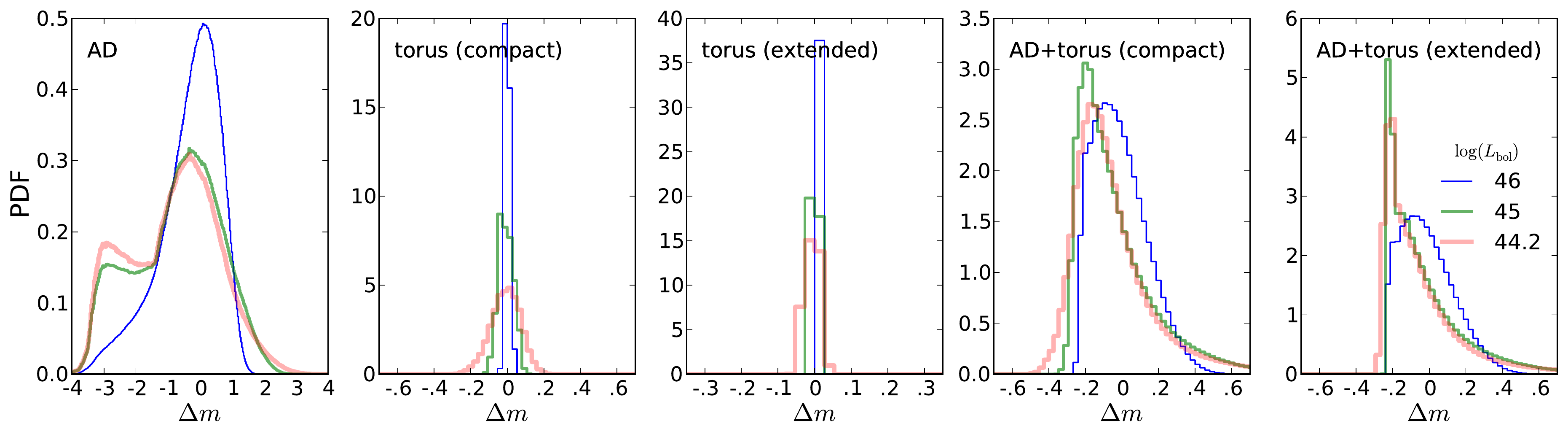}}
\setcounter{subfigure}{1}
\subfigure[Minimum image]{\includegraphics[scale=0.45]{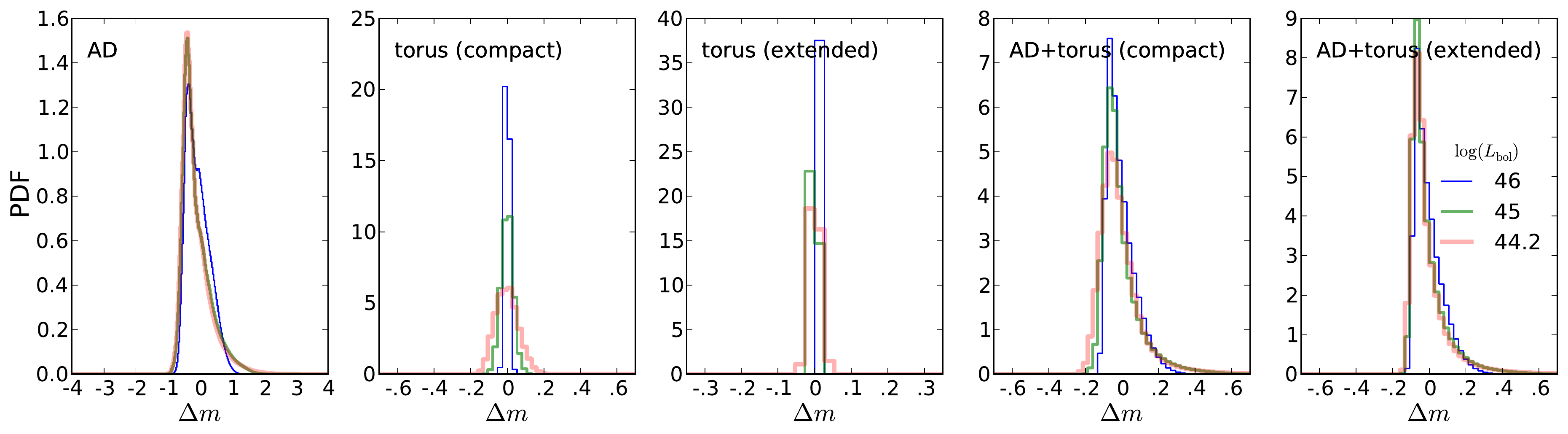}}
\caption{Probability density function (PDF) for microlensing $\Delta m$ at 2.2 $\micron$ for the accretion disc (AD), the torus (compact and extended model) and compound model (torus+accretion disc). The distributions are given for three different luminosities $\log(L_{\rm {bol}} / {\rm{erg/s}})$=44.2, 45, 46. Note that in our convention $\Delta m > 0$ corresponds to magnification.}
\label{fig:compaL}
\end{figure*}

We first investigate how the luminosity of the source and the compactness of the torus modify the probability of microlensing. We follow  the prescriptions of Sect.~\ref{sec:source} for $\lambda = 2.2\,\micron$ to calculate the brightness profile of the source. In order to gain intuition on the relative effect of the dust and of the accretion disc, we show in Fig.~\ref{fig:compaL} the PDF of microlensing for each component of the emission separately. The three left-most panels show the distribution for the accretion disc, for a compact torus and for an extended torus, while the two right-most panels show the distribution for the compound source models. The accretion disc contributes to 20\% of the compound source. In each panel, the PDF for three different source luminosities $\log(L_{\rm {bol}}/{\rm {erg/s}}) = [46, 45, 44.2]$ are compared. Panel (a) shows the PDF for a saddle-point image, and panel (b) for a minimum image. In both cases, we fix the fraction of compact matter to $\kappa_* =$\,0.07. We compare panels (a) and (b) in Sect.~\ref{subsec:parity} while here we focus on the results of the saddle-point image (i.e. panel (a)).

First, this figure confirms that the compact emission from the accretion disc can be significantly microlensed for all luminosities considered. Second, we see that if the torus is as compact as suggested by near-infrared interferometric data, the probability for it to be significantly microlensed is non-negligible, even for intrinsically bright sources. On the other hand, if the torus emission is more extended, the probability of getting the torus microlensed by $> 0.1$\,mag drops to almost zero. Finally, we see from the rightmost panels that the small contribution of the flux from the accretion disc governs the probability distribution of microlensing, such that independently of the torus model, there is always a significant probability for the source to be microlensed in that wavelength range.  

Another interesting feature unveiled by that figure is the existence of a ``cut-off'' in the PDF of the compound model for large demagnification (the PDF drops sharply at $\Delta m \sim -0.25$\,mag). This effect is discussed in Sect.~\ref{subsec:fAD}.

\subsection{Changing the image type}
\label{subsec:parity}

In this section we compare the PDF of microlensing for a saddle-point image (Fig.~\ref{fig:compaL}a) and a minimum (Fig.~\ref{fig:compaL}b). We can see that the accretion disc (left-most panel) emission in a saddle point image has a much larger probability to be strongly demagnified ($\Delta m < -1.1$\,mag) than the minimum image. This difference in behaviour between saddle-point and minimum is a well known effect and we encourage the interested reader to consult~\cite{Schechter2002} and ~\cite{Saha2011a} for a deeper understanding of this effect. Interestingly, we see that this difference becomes weaker with the increasing source size. This is particularly clear for the torus, when comparing the PDF for the saddle-point and for the minimum image. This similarity of the PDF between the minimum and the saddle-point persists for the compound model. The origin of this effect is the ``suppression'' of the large demagnification in saddle-point images. Indeed, when strong demagnification occurs, the flux of the 
accretion disc gets diluted in the flux of the torus to such an extent that only a weak demagnification can be observed. It is however important to note that the suppression of large demagnification in the near-infrared is only valid in the context of microlensing by a population of solar-mass-like microlenses. If the Einstein radius of a microlens is larger than the torus size, i.e. the situation encountered in case of milli-lensing by substructures in the lensing galaxy with $M > 10^5 M_{\sun}$, then large demagnification of the torus (and therefore of the compound model) can still occur.

\subsection{Changing the lens properties}
\label{subsec:lens}

\begin{figure*}
\centering

{\includegraphics[scale=0.45]{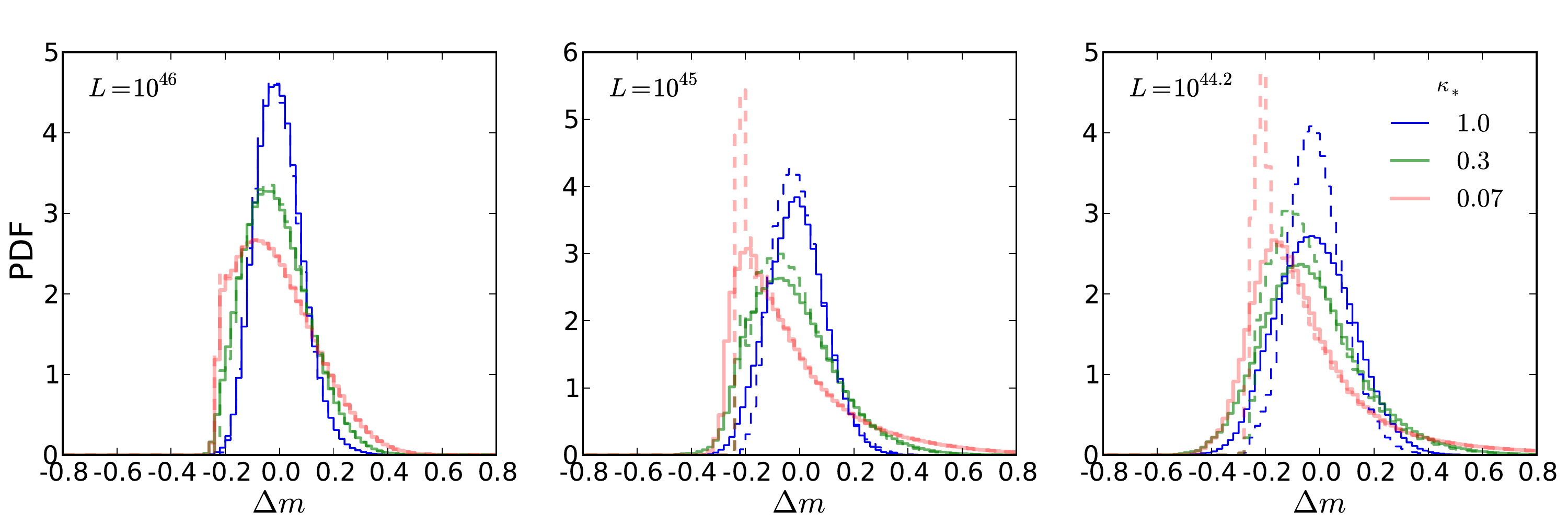}}

\caption{PDF for microlensing $\Delta m$ of a saddle point image at 2.2\,$\micron$ for three different fractions of compact matter ($\kappa_*$) and for three different luminosities $\log(L_{\rm {bol}} / {\rm{erg/s}})$=46, 45, 44.2. Solid lines correspond to the compact torus and dashed lines to an extended torus. Note that in our convention $\Delta m > 0$ corresponds to magnification.}
\label{fig:compakappas}
\end{figure*}

The fraction of compact matter in front of a strongly lensed image is in the range of 5 to 10\% \citep{Mediavilla2009, Pooley2012}, but seems to be several tens of percent in other systems or even reach 100\% for the Einstein Cross \citep{Dai2010a, Poindexter2010b}. In Fig.~\ref{fig:compakappas}, we show the PDF for microlensing of a compound source at $\lambda\,=\,$2.2\,$\micron$ for three different values of $\kappa_*$. The results are displayed for three different source luminosities, and for the two torus models. It can be seen that a decrease of $\kappa_*$ leads to a shift of the peak of the PDF towards significant demagnification. This is again a well known effect for saddle-point images \citep{Schechter2002}. However, it is important to notice that this effect is mostly caused by the sizable contribution of the accretion disc to the total flux (see Sect.~\ref{sec:discussion}). 

\subsection{Changing the wavelength and $f_{\rm{AD}}$}
\label{subsec:lambda}

\begin{figure*}
\centering
\includegraphics[scale=0.45]{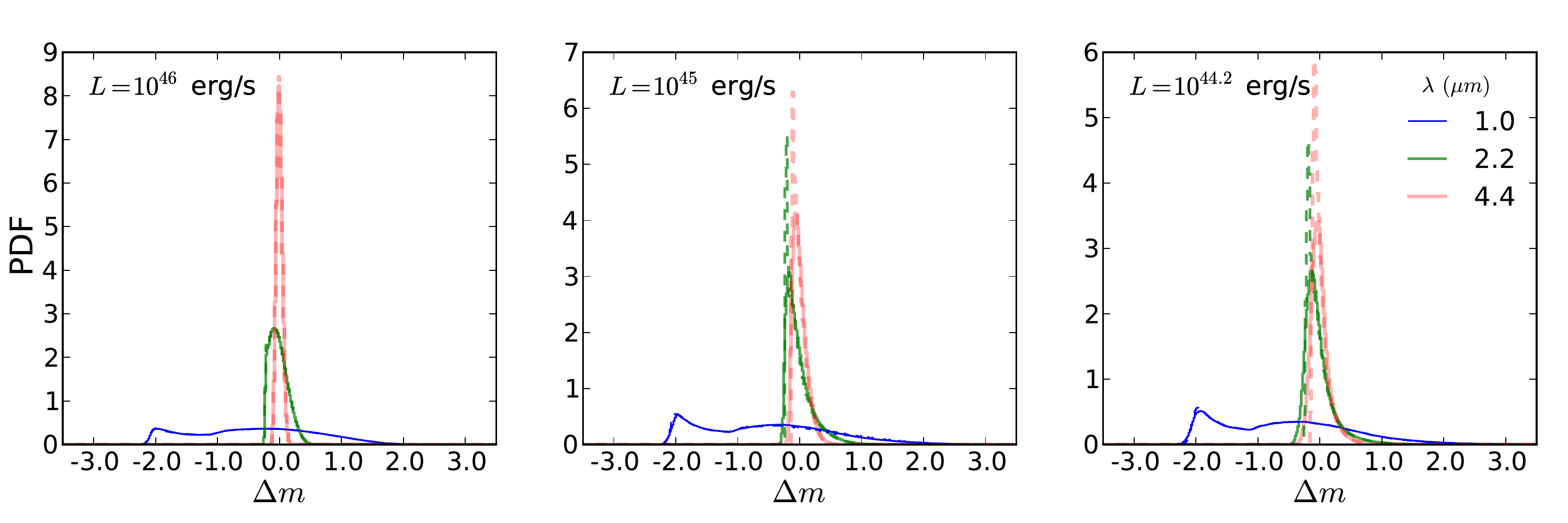}\\
\caption{PDF for microlensing $\Delta m$ of a saddle point image at three different wavelengths $\lambda =$\,1.,2.2,4.4\,$\micron$, and three different luminosities $\log(L_{\rm {bol}} / {\rm{erg/s}})$=46, 45, 44.2. Solid lines are for a compact torus and dashed lines for an extended torus. Note that in our convention $\Delta m > 0$ corresponds to magnification.}
\label{fig:compalambda}
\end{figure*}

The appearance of the source varies significantly from 1 to 4.4\,$\micron$. The main change comes from the decreasing contribution of the accretion disc emission to the total flux with increasing wavelength. According to our model, the decrease remains however relatively small, and 10\% of the flux still comes from the disc at 4.4\,$\micron$ (see Sect.~\ref{subsec:fAD}). On the other hand, the accretion disc size also increases with wavelength. We display in Fig.~\ref{fig:compalambda} the PDF for three wavelengths and luminosities. We see that the shape of the PDF changes with increasing wavelength (peaking at larger $\lambda$) because of the progressive dilution of the flux of the accretion disc in the flux of the torus, not because of the increase in the size of the torus and of the disc with $\lambda$.

\begin{figure*}

\centering
\begin{tabular}{lll}
\includegraphics[scale=0.29]{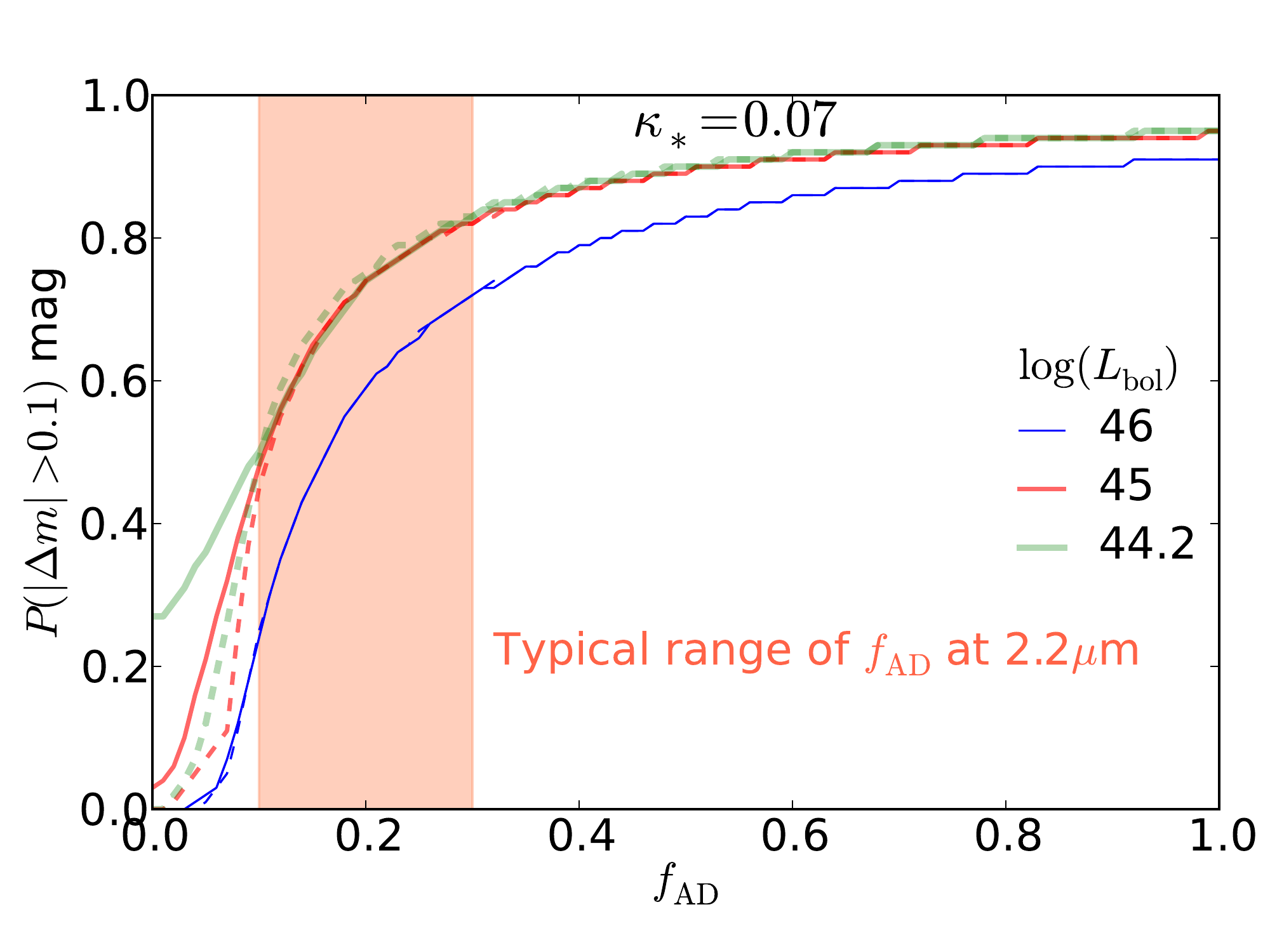} & \includegraphics[scale=0.29]{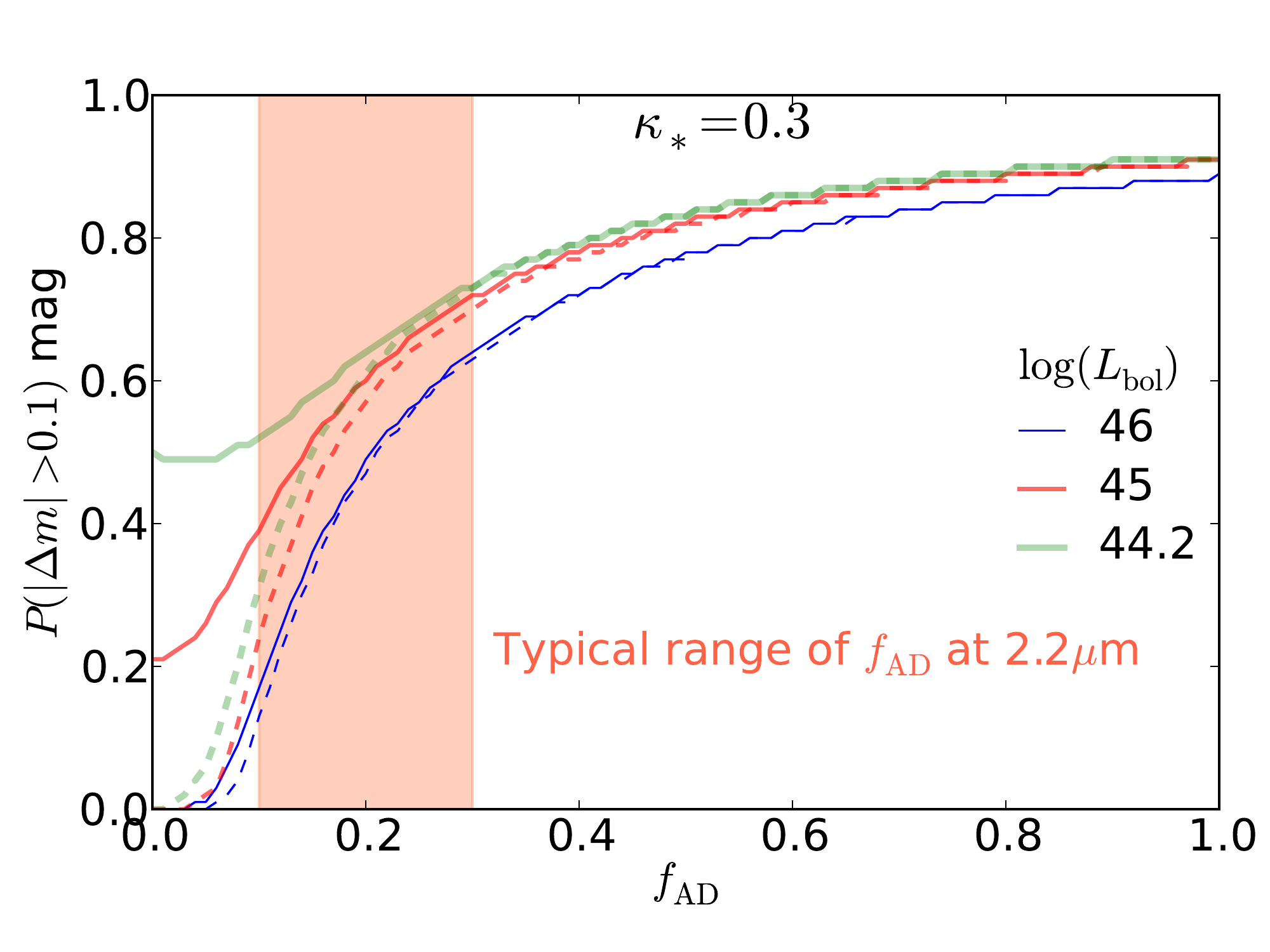}& \includegraphics[scale=0.29]{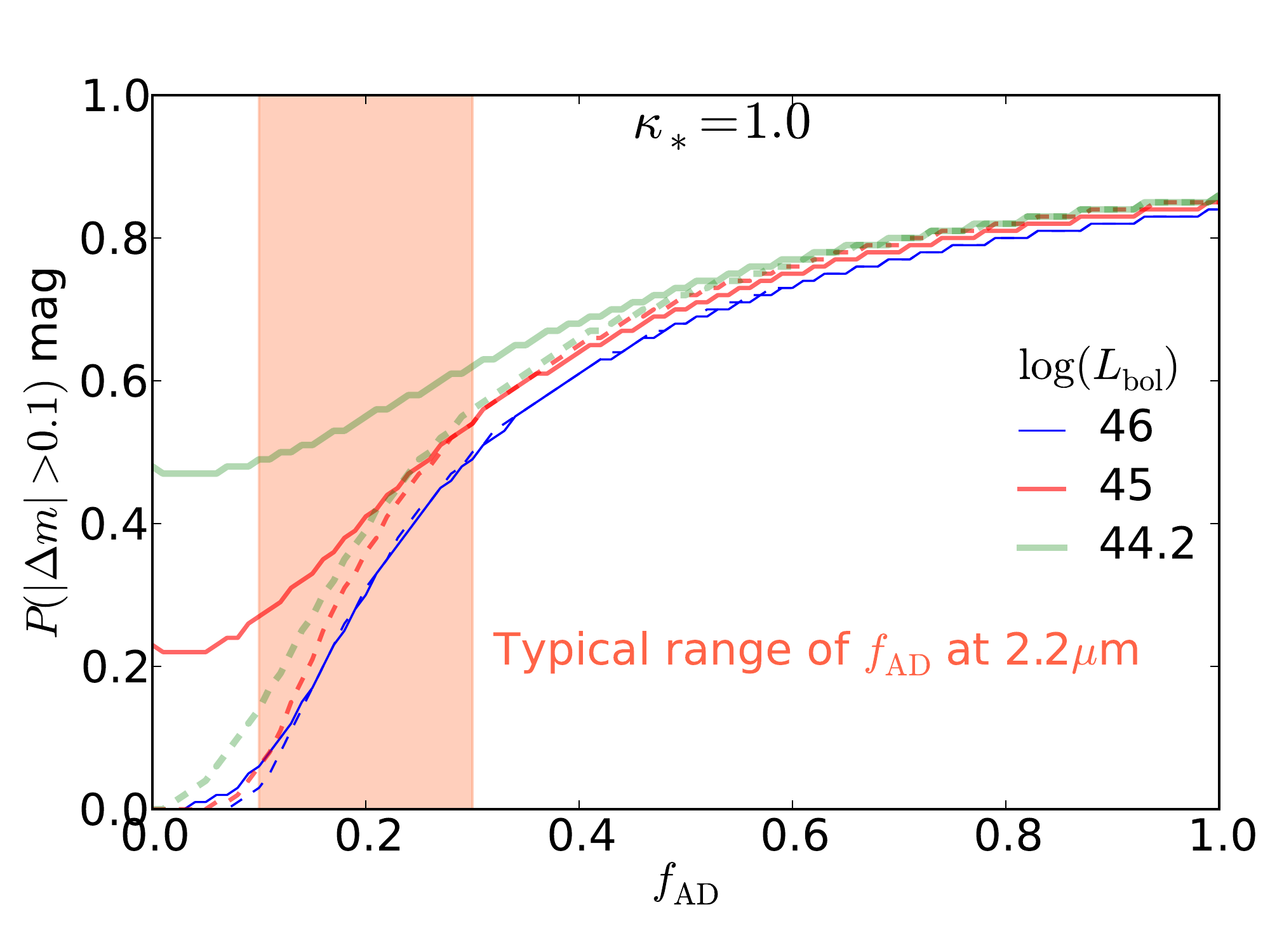}\\
\end{tabular}
\caption{Probability to observe microlensing (de)magnification $|\Delta m| > 0.1$ mag as a function of the fraction of the flux $f_{\rm{AD}}$ originating from the accretion disc. The solid line is for a compact model for the torus and the dashed line for an extended model. The blue, red and green curves correspond to different source bolometric luminosities. Panels from left to right correspond to $\kappa_*=0.07$,  $\kappa_*=0.3$, $\kappa_*=1.0$. Probabilities are calculated for $\lambda_{\rm{rest}} = 2.2\,\micron$. }
\label{fig:fADvar}
\end{figure*}

Since we see that the accretion disc is the main cause of microlensing in the rest-NIR, we investigate how the probability of microlensing changes with $f_{\rm{AD}}$. For this purpose we fix the accretion disc and torus sizes to their values at 2.2\,$\micron$, and calculate the probability of observing microlensing (de)magnification $|\Delta m| > 0.1$ mag for different values of $f_{\rm{AD}}$. Figure~\ref{fig:fADvar} shows the results for different values of $\kappa_*$. We see that the probability to observe $|\Delta m| > 0.1$ mag is already significant for $f_{\rm{AD}} \sim 0.05$. The compactness of the torus only weakly changes the results, except when $L_{\rm {bol}} < 10^{45}$\,erg/s and when $f_{\rm{AD}}$ is small. This is indeed the regime where the compact torus can be significantly microlensed such that the probability of microlensing gets boosted despite the small contribution of flux from the accretion disc.  

\subsection{Typical probabilities of microlensing}
\label{subsec:summary}

\begin{table*}[tb]
\caption{Probability of observing microlensing (de)magnification $|\Delta m| > 0.1$\,mag in a saddle-point (minimum) image, for three source luminosities $L$, and three values of $\kappa_*$. Probabilities are calculated for a pure accretion disc (AD) at $2.2\,\micron$, a pure torus (compact or extended), and for compound source models as they would be observed at $\lambda=2.2\,\mu$m, 1.0\,$\mu$m, 4.4\,$\mu$m. 
} 

\begin{center}
\begin{tabular}{lccc|ccc|ccc}
\hline 

\multicolumn{10}{c}{Saddle-point image} \\
\hline
 & \multicolumn{3}{c|}{$L=10^{46}$\,erg/s} & \multicolumn{3}{|c|}{$L=10^{45}$\,erg/s} & \multicolumn{3}{|c}{$L=10^{44.2}$\,erg/s} \\ 
\hline
& \multicolumn{3}{c|}{$\kappa_*$} & \multicolumn{3}{|c|}{$\kappa_*$} & \multicolumn{3}{|c}{$\kappa_*$} \\

               & 0.07 & 0.3 & 1.0 &  0.07 & 0.3 & 1.0 & 0.07 & 0.3 & 1.0  \\
 \hline 
AD             &0.91 & 0.89 & 0.84 & 0.95 & 0.91 & 0.85 & 0.95 & 0.91 & 0.86  \\
\hline
Compact        &0.00 & 0.00 & 0.00 & 0.03 & 0.21 & 0.23 & 0.27 & 0.50 & 0.48  \\
$1.0\,\micron$ &0.94 & 0.89 & 0.83 & 0.94 & 0.90 & 0.84 & 0.94 & 0.91 & 0.84  \\
$2.2\,\micron$ &0.59 & 0.49 & 0.30 & 0.74 & 0.60 & 0.41 & 0.74 & 0.64 & 0.55  \\
$4.4\,\micron$ &0.06 & 0.12 & 0.07 & 0.48 & 0.41 & 0.27 & 0.54 & 0.53 & 0.49  \\
\hline
Extended       &0.00 & 0.00 & 0.00 & 0.00 & 0.00 & 0.00 & 0.00 & 0.00 & 0.00  \\
$1.0\,\micron$ &0.94 & 0.89 & 0.83 & 0.94 & 0.90 & 0.84 & 0.94 & 0.91 & 0.84  \\
$2.2\,\micron$ &0.59 & 0.47 & 0.31 & 0.74 & 0.57 & 0.36 & 0.75 & 0.61 & 0.39  \\
$4.4\,\micron$ &0.05 & 0.08 & 0.04 & 0.46 & 0.27 & 0.08 & 0.56 & 0.36 & 0.17  \\

\hline
 \multicolumn{10}{c}{Minimum image} \\
\hline

AD             &0.84 & 0.87 & 0.86 & 0.89 & 0.89 & 0.88 & 0.89 & 0.90 & 0.88\\
\hline                                                                        
Compact        &0.00 & 0.00 & 0.02 & 0.01 & 0.25 & 0.35 & 0.17 & 0.48 & 0.57\\
$1.0\,\micron$ &0.86 & 0.88 & 0.86 & 0.87 & 0.89 & 0.87 & 0.87 & 0.89 & 0.87\\
$2.2\,\micron$ &0.23 & 0.42 & 0.39 & 0.35 & 0.53 & 0.54 & 0.41 & 0.60 & 0.59\\
$4.4\,\micron$ &0.00 & 0.08 & 0.14 & 0.13 & 0.37 & 0.43 & 0.27 & 0.51 & 0.57\\
\hline                                                                        
Extended       &0.00 & 0.00 & 0.00 & 0.00 & 0.00 & 0.00 & 0.00 & 0.01 & 0.04\\
$1.0\,\micron$ &0.86 & 0.88 & 0.86 & 0.87 & 0.89 & 0.87 & 0.87 & 0.89 & 0.87\\
$2.2\,\micron$ &0.21 & 0.40 & 0.37 & 0.30 & 0.51 & 0.43 & 0.33 & 0.53 & 0.47\\
$4.4\,\micron$ &0.00 & 0.04 & 0.06 & 0.06 & 0.17 & 0.12 & 0.08 & 0.30 & 0.28\\

\hline
\end{tabular} 
\label{tab:Pall}
\end{center} 
\end{table*}

Table~\ref{tab:Pall} summarises our results. We provide the probability for a source to be microlensed by more than 0.1~\,mag for different source luminosities and fraction of compact objects $\kappa_*$ in the lens. We give those probabilities for the accretion disc (AD) at $2.2\,\micron$, for our two different kinds of tori (i.e. compact and extended torus, see Table ~\ref{tab:size}), and for the compound source model (AD+torus) at 3 different rest-frame wavelengths.  

Table~\ref{tab:Pall} shows that microlensing of the dust torus amounts more than 0.1 mag for the compact torus model and sources with $L_{\rm {bol}}\leq 10^{45}$ erg/s. Microlensing of the torus remains below 0.1 mag for the extended torus. The probability of significant microlensing of the (compact) torus increases with $\kappa_*$, conversely to the behaviour observed for the accretion disc. These results are qualitatively independent of the image type (i.e. minimum or saddle-point). 

The probability of microlensing for the compound model is governed by the relative contribution of the accretion disc flux to the total flux. As shown in Fig.~\ref{fig:fADvar}, values of $f_{\rm{AD}} \sim 0.05$ are sufficient to enable microlensing $\Delta m > 0.1$\,mag. As soon as the accretion disc contribution is large enough, the probability of microlensing depends on both the image type and $\kappa_*$. We see at e.g. $4.4\,\micron$ that saddle-point images show a larger probability of microlensing $\Delta m > 0.1$ mag for small $\kappa_*$, while minimum images show a large probability for $\kappa_*\,=\,$1.

\section{Discussion}
\label{sec:discussion}

The non-negligible probability of observing a microlensing signal of at least 0.1 mag in the rest-frame range 1--4.4\,$\micron$, implies that microlensing cannot be neglected in the study of lensed quasars flux ratios from $K-$band up to 11\,$\micron$ (observed frame). We discuss hereafter how to link our simulations to observed flux ratios in lensed quasars. 

\subsection{Contribution of the accretion disc}
\label{subsec:fAD}

\begin{figure}
\centering
\includegraphics[scale=0.45]{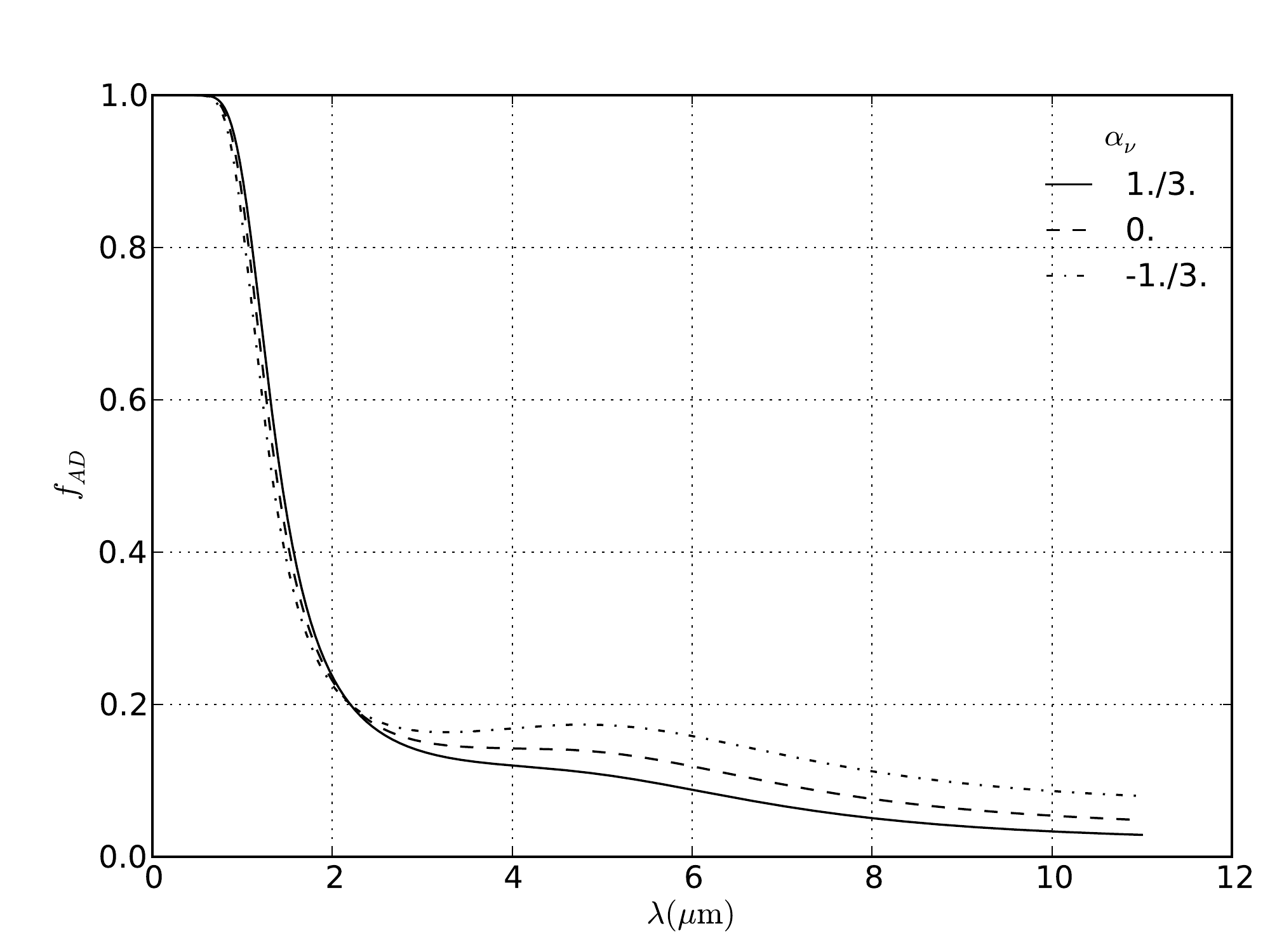}
\caption{Variation of the fraction of the flux coming from the accretion disc ($f_{\rm{AD}}$) with rest-frame wavelength. $f_{\rm{AD}} (\lambda)$ is shown for three different power law index $\alpha_\nu$ of the accretion disc emission. For the three curves we have fixed $f_{\rm{AD}} (2.2\,\micron)$ = 0.2, the fiducial value used in this work. }
\label{fig:fADlambda}
\end{figure}

The fraction of the flux originating from the accretion disc ($f_{\rm{AD}}$) controls the amount of microlensing in the NIR-MIR. In our model, we used a fiducial value $f_{\rm{AD}}\,=\,0.2$ at $\lambda_{\rm {rest}}\,=\,$2.2\,$\micron$. This value is typical of what is observed in local AGNs (i.e. $f_{\rm{AD}}$ in the range 14--29\%). We show in Fig.~\ref{fig:fADlambda} how $f_{\rm{AD}}$ varies with wavelength. It is important to realise that $f_{\rm{AD}} (\lambda)$ may change depending of the real slope of the power law accretion disc emission in the near-infrared. In our model, we used a standard index $\alpha_\nu=1/3$. This choice is supported by spectropolarimetry of a sample of local AGNs \citep{Kishimoto2008}. It contrasts with the redder spectral slope $\alpha_\nu \sim -0.4$ observed in the optical-ultraviolet domain \citep[e.g.][]{Vandenberk2001}. Such a redder slope, if valid at larger wavelengths, would lead to an even larger contribution from the disc in the NIR-MIR and would increase the contribution of microlensing in that 
range. We illustrate the effect on $f_{\rm{AD}} (\lambda)$ for different values of $\alpha_\nu$ in Fig.~\ref{fig:fADlambda}.

The various PDF shown in Figs.~\ref{fig:compaL},~\ref{fig:compakappas},~\ref{fig:compalambda} display a sharp cutoff for demagnification but not for magnification. The origin of this cut-off is caused by the effective increase of contrast between the flux from the accretion disc and from the torus in case of large demagnification. When this situation occurs, the effective fraction of the flux coming from the accretion disc drops significantly, such that the total flux gets dominated by emission from the torus and that extreme demagnification has no effect anymore. In case of micro-magnification, the inverse occurs, and the total flux becomes larger than if the torus was the only source of emission. 

The cut-off value of the PDF can be derived by expressing the microlensing of the compound model: 
\begin{equation}
\Delta m = 2.5\,\log \left[ f_{\rm{AD}}\,\mu_{\rm {AD}} + \mu_{\rm{torus}}\,(1-f_{\rm{AD}}) \right ].
\label{equ:compound}
\end{equation}

\noindent From this equation, we see that when the accretion disc is significantly de-magnified, $\mu_{\rm {AD}} \rightarrow 0$, we have $\Delta m \rightarrow 2.5\,\log [\mu_{\rm{torus}}\,(1-f_{\rm{AD}})]$. Therefore, when the demagnification of the torus remains small ($\mu_{\rm{torus}} \rightarrow 1$), the minimum demagnification of the source will be $\Delta m =2.5\,\log (1-f_{\rm{AD}})$. On the other hand, there is no formal upper limit to the maximum magnification, and the latter is only governed by the maximum amount of microlensing of the accretion disc \citep[see][]{Refsdal1991, Refsdal1997}, and by $f_{\rm{AD}}$. 

\subsection{Extended vs compact torus}

We find that noticeable microlensing of the torus is likely to take place with amplitudes as large as 0.15-0.2\,mag for objects with $L_{\rm{bol}} \sim 10^{44.2}$\,erg/s if a compact model for the torus is assumed. Based on interferometric studies of nearby AGNs, this model is appropriate for rest-frame wavelengths $\sim$ 2.2\,$\micron$. At a rest-frame wavelength of 5.5\,$\micron$ or larger\footnote{There is unfortunately no measurement in the rest-frame range 2.2--8.5\,$\micron$.}, the hot component of the torus gets several times more extended than at 2.2\,$\micron$, especially for  $L_{\rm bol} \leq 10^{45}$\,erg/s \citep[see Fig.~7 of ][]{Kishimoto2011b}. Therefore, the simulation considering an extended torus may be more appropriate when dealing with $\lambda \sim 4.4\,\micron$ (rest-frame). For larger luminosities ($L_{\rm bol} \sim 10^{46}$\,erg/s), the torus remains compact ($R_{\rm{out}}/R_{\rm{in}} \sim 2$), but the half-light radius is significantly larger than a microlens Einstein radius such that 
microlensing 
does not exceed 0.05\,mag (Fig.~\ref{fig:compaL}). Consequently, it is correct to consider that at 4.4\,$\micron$ (rest-frame), most of the microlensing originates from the disc. This is however with the assumption that there is no evolution with time of the torus properties. This is supported by a recent study of the properties of the 1-10\,$\micron$ SED of AGNs detected in the COSMOS field which finds no strong evolution of the SED with redshift \citep{Hao2012}.

\subsection{The effect of inclination}
\label{subsec:PA}

\begin{figure*}
\centering
\includegraphics[scale=0.45]{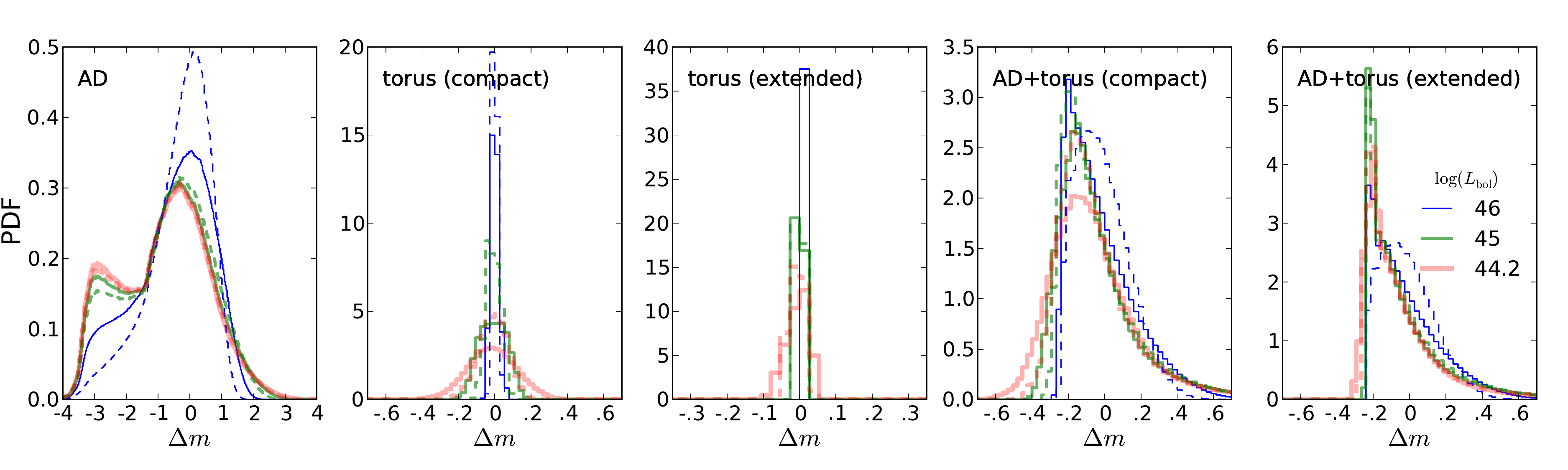}
\caption{Effect of inclination on the probability density function (PDF) of microlensing of a saddle-point image. The dashed lines show the PDF of Fig.~\ref{fig:compaL} (face-one quasar) and the solid lines show the corresponding PDF when the polar axis of the quasar is inclined by 60\,$\deg$. The distributions are given for $\log(L_{\rm {bol}} / {\rm{erg/s}})$=44.2, 45, 46. Note that in our convention $\Delta m > 0$ corresponds to magnification.}
\label{fig:compaL60}
\end{figure*}

Type 1 AGNs are not always face-on as considered in our model and inclinations of the polar-axis with respect to our line-of-sight as large as 60\,$\deg$ are plausible. Inclination will reduce the observed area covered by the source and thus increase the probability of important microlensing. We show in Fig.~\ref{fig:compaL60} the PDF for a saddle-point image at $\lambda_{\rm{rest}} = 2.2\,\micron$, and an inclination of the polar axis of $60\,\deg$. We see that the microlensing distribution of the torus gets broadened by a factor of about two. On the other hand the probability of large (de)-magnification of the accretion disc gets slightly increased, especially for the intrinsically brighter sources. These two effects lead to a broadening of the PDF of the compound model and to a shift of the peak of PDF towards demagnification for $L_{\rm {bol}}=10^{46}$\,erg/s. 

We have also examined the effect of inclination for a minimum image and $\kappa_*=0.07$. In that case, the distribution of magnification of the accretion disc remains roughly unchanged, but the magnification distribution of the torus gets broader, similarly to the situation encountered for the saddle-point. However, this effect remains too small to significantly modify the PDF of the compound model associated to the minimum. 

\subsection{Existing observations and simulations}

To our knowledge, there are only a few observational studies of the SED of lensed quasars in the NIR-MIR range. \cite{Agol2000} were the first to study the MIR flux ratios in the Einstein cross $\equiv$ Q2237+0305. They aimed to disentangle the synchrotron and dust emission origins of infrared AGN flux. Their analysis demonstrated that the MIR emission (observed frame) in Q2237+0305 originates from a region larger than one microlensing Einstein radius $\eta_0$ (for $\avgg{M} = 0.3\,M_{\sun}$), ruling out pure synchrotron emission. This result was later confirmed by \cite{Wyithe2002} who explored a larger family of macro-lens models. Recently, \cite{Agol2009} used updated MIR flux ratios obtained with the IRAC camera on Spitzer to improve the study of the SED of Q2237+0305. Their observations (see their Fig.~7) showed a clear chromaticity of the flux ratios between 0.6 and 11\,$\micron$ (observed frame). A similar trend was observed by \cite{Minezaki2009} using ground-based MIR-data. \cite{Agol2009} demonstrated that in order to reproduce the chromaticity of the flux ratios, the emission in the rest-range 1--4.2$\,\micron$ had to originate from two regions, a hot torus and an accretion disc with a temperature profile compatible with prediction from the standard thin accretion disc model. Two other quadruply lensed quasars have been observed in the MIR by \cite{Chiba2005}. These authors, searching for milli-lensing in these systems, also suggest significant compact emission in the system B1422+231, but not in PG1115+080. The doubly-lensed quasar HE1104$-$1805 has also been observed in the MIR and its flux ratios in the context of microlensing have been analysed by \cite{Poindexter2008}. The lack of chromaticity of the flux ratios in that system suggests that the emission down to $\sim 2.2\,\micron$ was dominated by an extended emission possibly originating from the torus. The data however do not rule out that a minor fraction of that emission comes from the disc. Observations of lensed systems above 2.2\,$\micron$ are generally sparse. Chromaticity of the flux ratios in the NIR (up to 3.8$\,\micron$ observed frame) has been investigated in six lensed quasars by \cite{Fadely2011a}. Some systems show a ``plateau'' in the flux ratios between 2.2 and 3.8 $\micron$ while others show clear differences likely caused by microlensing or milli-lensing \citep{Fadely2011b, Sluse2012b}.

A recent paper of \cite{Stalevski2012}, which appeared after the start of this work, investigated the generic possibility of microlensing of the dust torus. Our work differs in several respects from \cite{Stalevski2012} and complements their results. The main conceptual difference between the two studies relates to the model of the source. The work from \cite{Stalevski2012} focuses on the torus while we also address the effect of the accretion disc on the flux ratio. On the other hand, they use a torus model derived from 3-D radiative transfer simulations of a two-phase medium \citep{Stalevski2012a, Baes2011, Baes2003b}. Our model is simpler, but it is based on observational measurements of the torus size. Owing to the use of $R_{\rm{in}}$ equal to Barvainis sublimation radius \citep{Barvainis1987}, they effectively use a torus with an inner radius 3 times larger than observed in nearby objects \citep{Suganama2006}, and than our model, for the same $L$. For these reasons, direct comparison of results have 
to be careful. Because of the difference of the definition of inner radius, their source with $L=10^{12}\,L_{\sun}=3.9\times10^{45}$\,erg/s should match our source with $L_{\rm {bol}}\sim 10^{46}$\,erg/s. Contrary to \cite{Stalevski2012}, we do not have a model accounting for the increase of $R_{out}$ with wavelength but we consider two extreme models for the torus, a compact and an extended torus. At $\lambda=2.2\,\micron$, their model compares to our compact torus model, and at $\lambda=4.4\,\micron$, to our extended torus model. From their Fig.~4 (left panel), for a lens\footnote{This choice of redshift is needed to roughly match the Einstein radius with our fiducial value of $\eta_0$.} at $z=0.5$, a maximum microlensing magnification of 0.06\,mag at 2.2\,$\micron$ is expected, in agreement with our Fig.~\ref{fig:compaL}. We cannot perform the comparison at 4.4$\,\micron$ because we assumed no microlensing of the extended torus for the high luminosity source.

\subsection{Variations with time}
\label{subsec:timescale}

\begin{figure}
\centering
\includegraphics[scale=0.48]{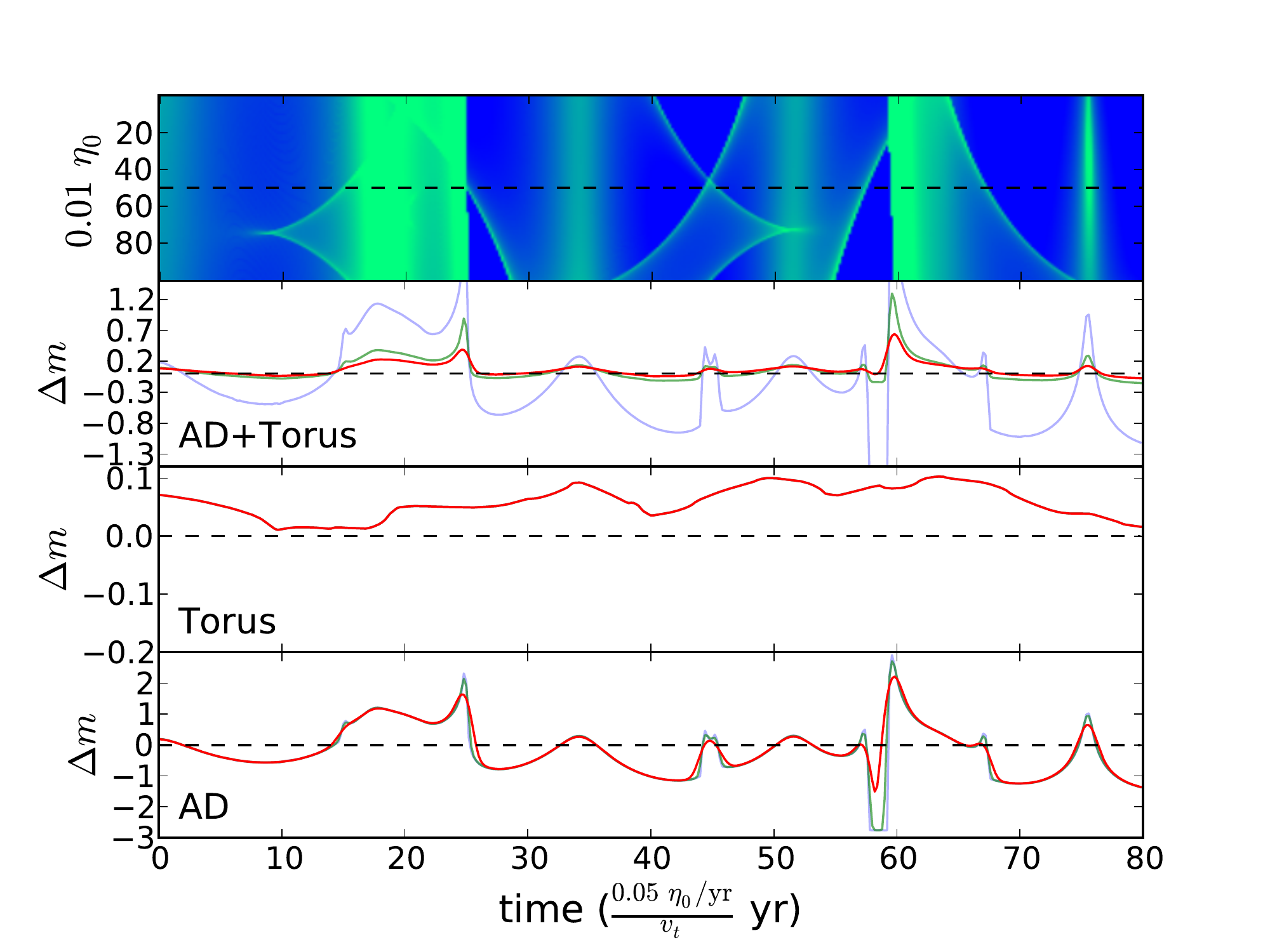}
\caption{Example of microlensing light curve for a saddle-point image (with $\kappa_* = 0.07$). The microlensing-induced flux variations (in mag) as a function of time are displayed for a source with $L_{\rm{bol}}=10^{44.2}$ erg/s, and a compact torus model. The microlensing event duration scales with the inverse of the net transverse velocity $v_t$ expressed in Einstein radius/year. The upper panel shows a portion of the microlensing map from where the light curve is extracted. Blue (dark) region correspond to demagnification and green (light) region to magnification. The other panels show the light curve for the accretion disc (AD), for the torus, and for the composite source (AD+torus). The results are presented for three different wavelengths, 1\,$\micron$ (light blue), 2.2\,$\micron$ (light green), and 4.4\,$\micron$ (red). Note that the torus model is the same for each wavelength.}
\label{fig:tvar}
\end{figure}

Contrary to milli-lensing, the amplitude of microlensing is expected to vary on a human time scale \citep{Schechter2002}. This time scale depends on the relative transverse velocity between the source, the microlens and the observer \citep{Kochanek2004a}. Typical source plane transverse velocities are $v_t \sim 0.05\,\eta_0/{\rm{yr}}$, and can reach $v_t \sim 0.15\,\eta_0/{\rm{yr}}$ in the most favourable cases \citep{Mosquera2011}. We show in Fig.~\ref{fig:tvar} an example of an expected microlensing light curve in a saddle-point image for a source with $L_{\rm {bol}} = 10^{44.2}$\,erg/s, and a compact torus. The figure shows the light curve separately for an accretion disc alone (AD), a torus model alone, and for a more realistic composite model (see Sect.~\ref{subsec:fAD}). The light curves are displayed for three different wavelengths: 1.1\,$\micron$, 2.2\,$\micron$, and 4.4\,$\micron$. It can be seen that the most drastic variations with wavelength are generally associated to caustic crossing-like events. In such cases, chromatic variations will be observed from the ultra-violet, where the accretion disc dominates, up to the near/mid-infrared range. If the source is located farther away from a caustic, then the chromaticity will be mostly associated to the variable fraction of flux from the accretion disc to the total flux. Another interesting feature is the suppression of large demagnifications when the flux of the torus takes over the flux of the accretion disc (cf Eq.~\ref{equ:compound}). Because the microlensing of the accretion disc does not show important chromaticity in this situation, we could in principle use large demagnification events to measure $f_{\rm{AD}}(\lambda)$. 

As outlined previously, the microlensing signal from the torus is relatively weak. In the most favourable case of our compact torus model, flux variations might be observed with amplitude as large as 0.1 mag on time scales of typically years to decade(s). For a brighter source and/or for more extended torus, the microlensing of the torus rarely reaches more than 0.05 mag and time scales for significant variations reach several decades.  

\section{Consequences for flux ratio anomalies}
\label{sec:fluxratios}

We investigate in this section the strength of the anomalies produced by microlensing in the MIR. To quantify this effect, we simulate microlensed flux ratios for minimum and saddle point images, and calculate the quantity $R_{\rm{cusp}}$. This quantity is commonly used to identify flux ratio anomalies caused by substructures \citep[e.g][]{Mao1998, Keeton2003, Xu2011}. The magnified fluxes $I_{A,B,C}$ of the bright saddle-point image A and two minima images B and C in a cusp-configuration (Fig.~\ref{fig:cusp}) can be used to calculate: 

\begin{equation}
R_{\rm cusp} = {\frac{{|{1-I_{\rm B}/I_{\rm A}-I_{\rm C}/I_{\rm A}}|}}{1+I_{\rm B}/I_{\rm A}+I_{\rm C}/I_{\rm A}}}. 
\label{equ:rcusp}
\end {equation}

\begin{figure*}
\centering
\setcounter{subfigure}{0}
\subfigure[Compact torus model]{\includegraphics[scale=0.45]{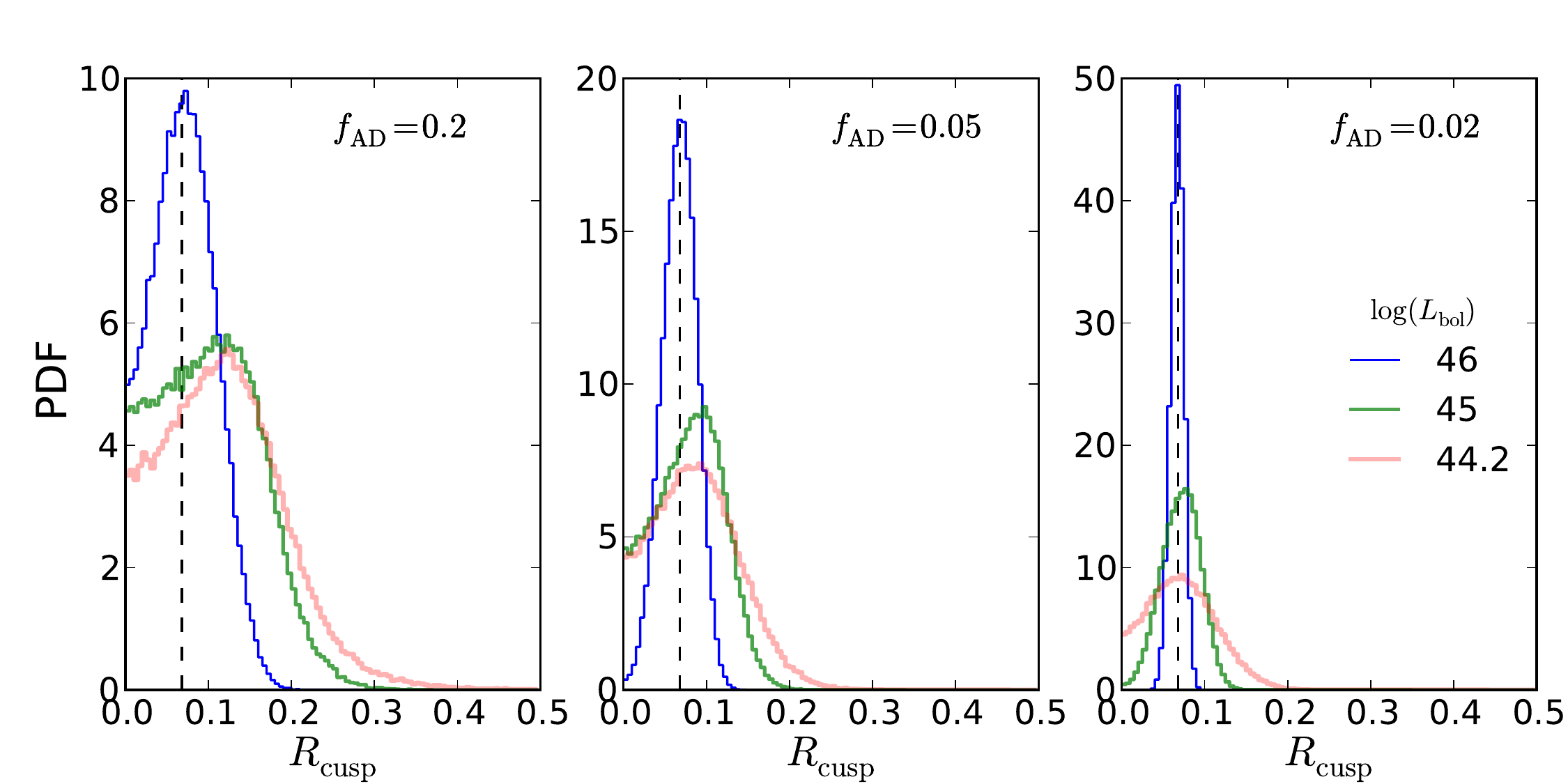}}
\setcounter{subfigure}{1}
\subfigure[Extended torus model]{\includegraphics[scale=0.45]{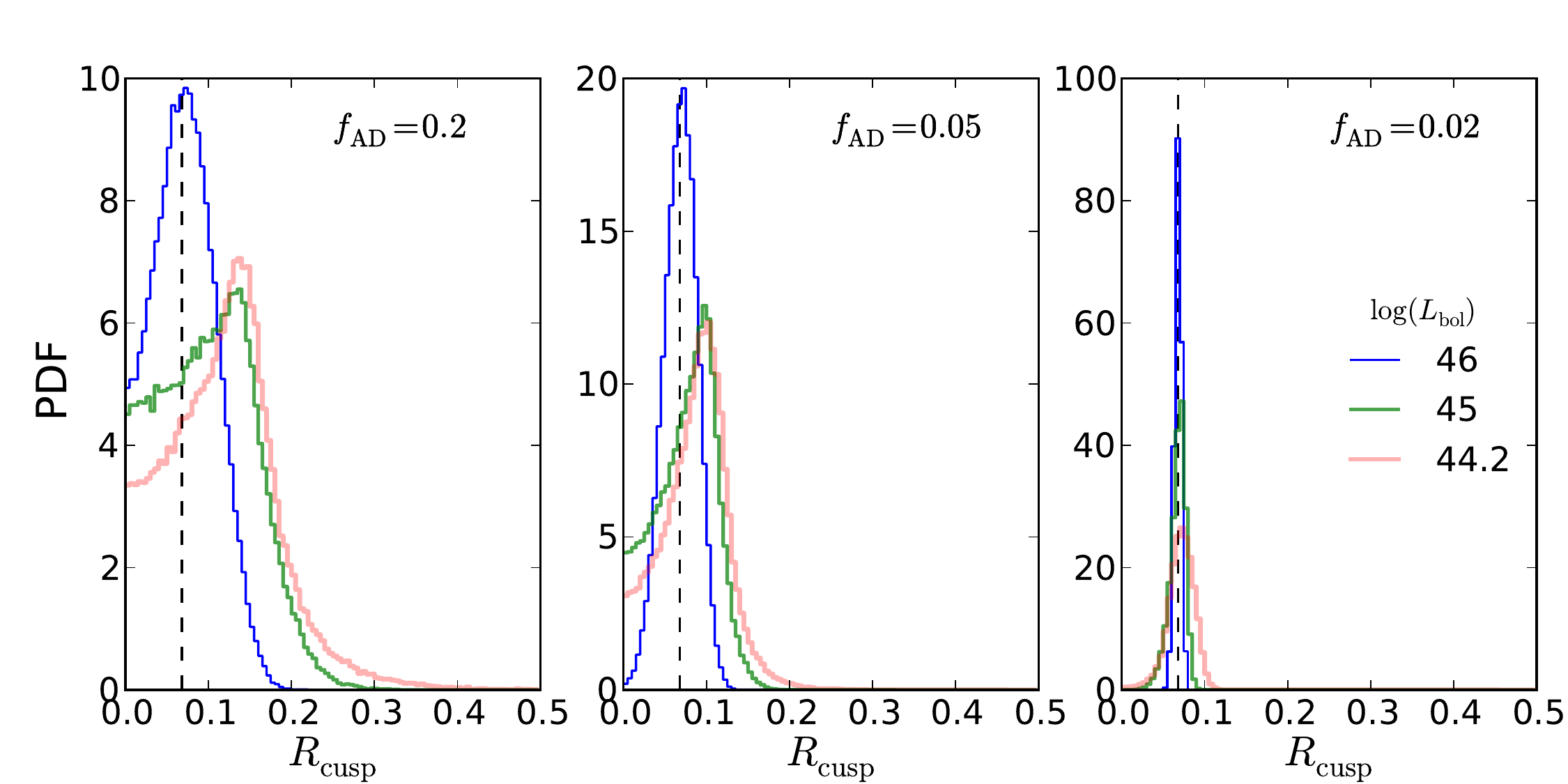}}
\caption{Probability density function $R_{\rm {cusp}}$ at 4.4\,$\micron$ (rest-frame) for $f_{\rm{AD}}=0.2$ (left), $f_{\rm{AD}}=0.1$ (middle), $f_{\rm{AD}}=0.02$ (right). The distributions are given for three different luminosities $\log(L_{\rm{bol}} / {\rm{erg/s}})$=44.2, 45, 46. The upper panel shows the distribution for a compact torus and the bottom panel for an extended torus. The value of $R_{\rm {cusp}}$ in absence of microlensing is $R_{\rm {cusp}} = 0.068$ and is depicted with a vertical dashed line. }
\label{fig:Rcusp}
\end{figure*}

\noindent In the ideal case, $R_{\rm {cusp}} \sim 0$. Significant deviations from zero or from the value estimated based on a simple macro model is considered as a good sign of an anomaly. For our fiducial values of  ($\kappa, \gamma$) of the saddle-point and minimum images, we calculate $R^{\rm{theo}}_{\rm {cusp}} = 0.068$. We show in Fig.~\ref{fig:Rcusp} the distribution of values of $R_{\rm{cusp}}$ in presence of microlensing, for different source luminosities, and for $f_{\rm{AD}} (\lambda = 4.4\,\micron) = $ 0.2, 0.1 and 0.02. We see that a large deviation from the expected value of $R_{\rm {cusp}}$ may be observed in the MIR range, especially for an intrinsically fainter system. 

\section{Consequences on the SED}
\label{sec:SED}

\begin{figure*}
\centering
\begin{tabular}{cc}
\includegraphics[scale=0.45]{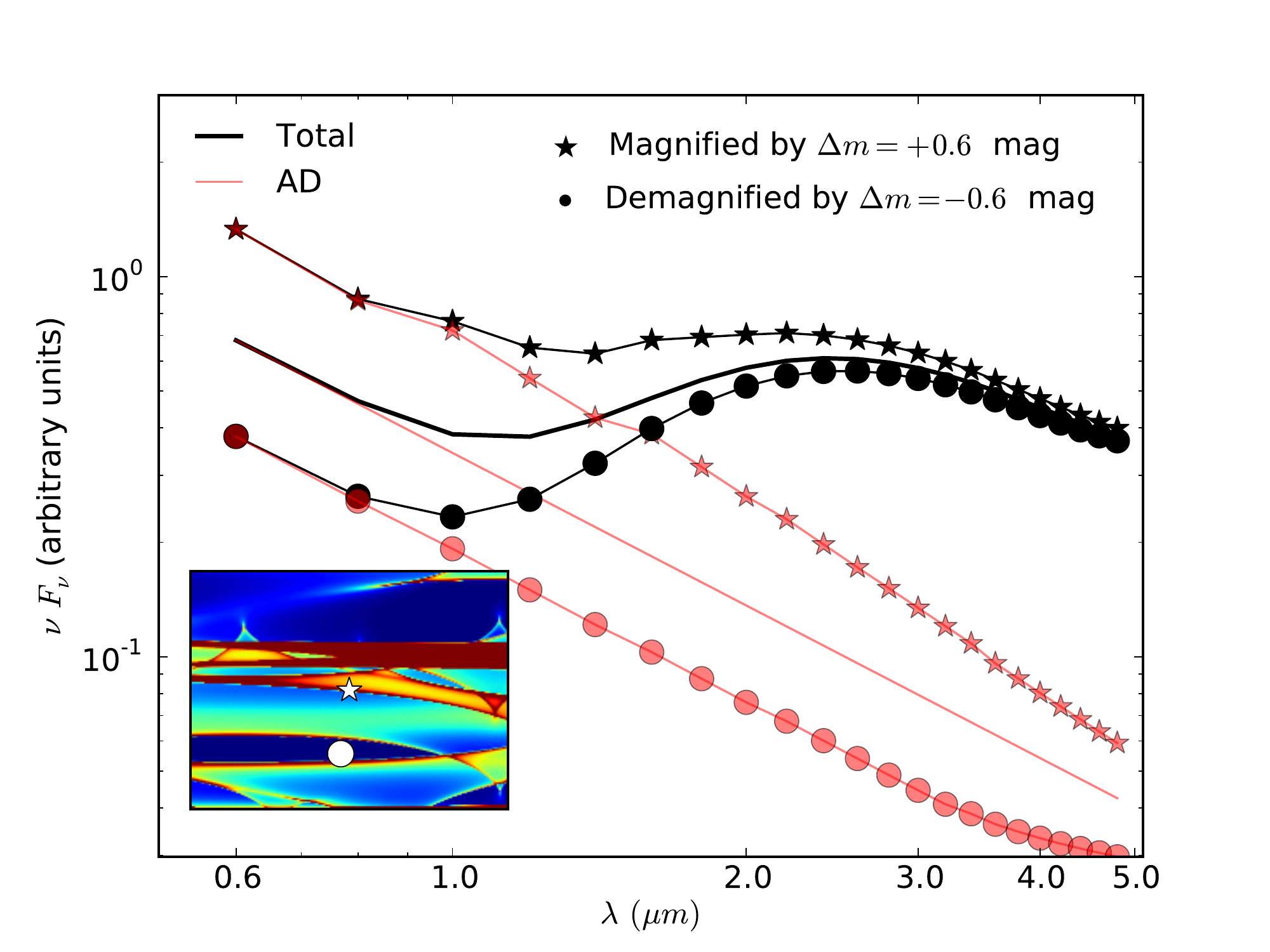} & \includegraphics[scale=0.45]{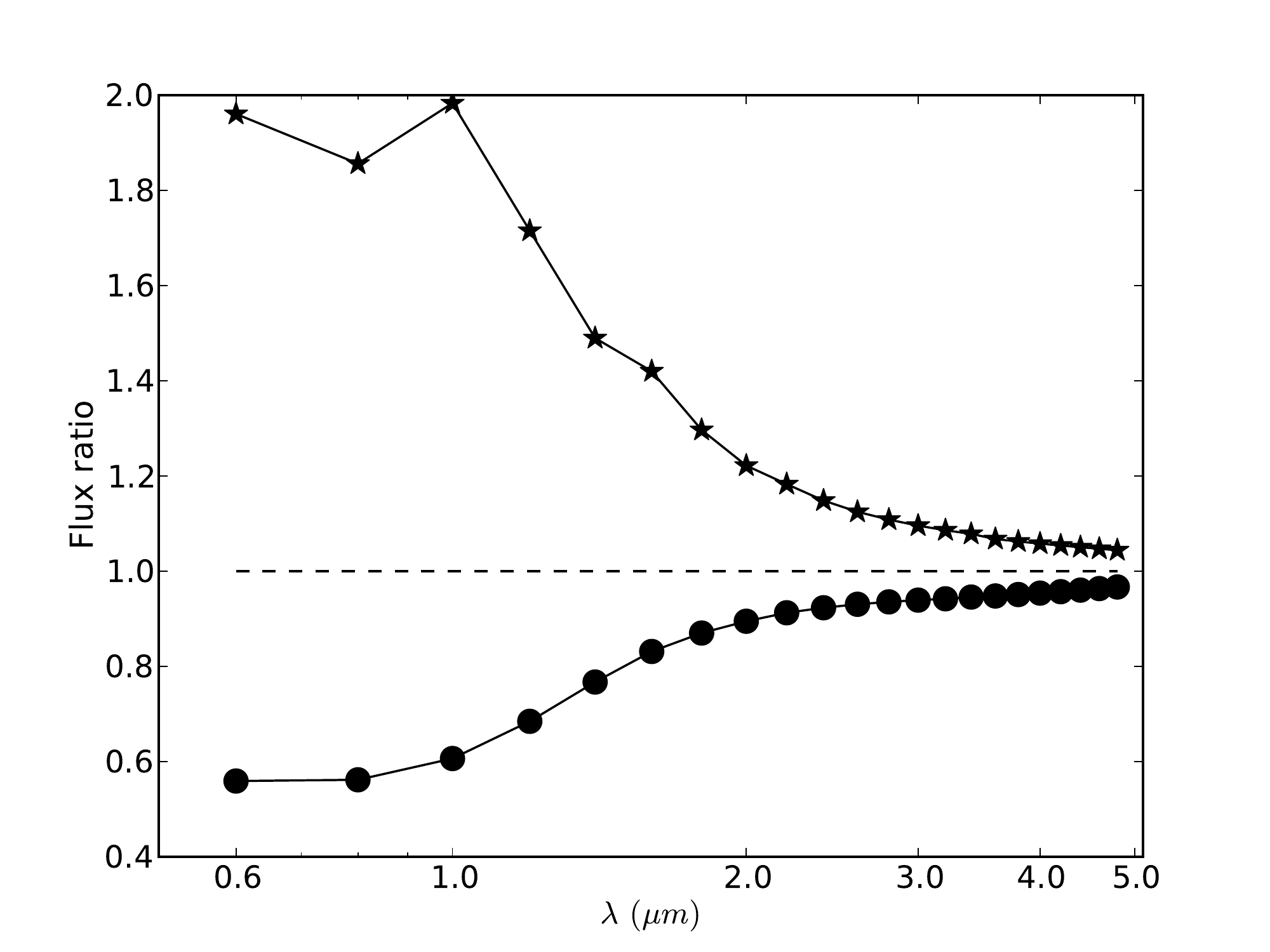} \\
\end{tabular}
\caption{{\it {Left}}: Example of modification of the SED due to microlensing. The reference SED of the accretion disc (red) and total SED (AD+torus; thick black) are shown with solid lines. The inset panel shows the source in two different regions of the microlensing map. The star shows the case of a micro-magnification of the AD by $\mu=1.84$ and the circle shows the case of a micro-demagnification of the AD by $\mu=0.57$ at 1.1$\micron$. The panel shows the corresponding SED of the accretion disc component (red) and of the total SED (black). {\it {Right}}: Corresponding ratio between the microlensed and the unlensed (total) SED. The bump in the upper curve illustrates the non linear behaviour of microlensing produced by the complex caustic network. }
\label{fig:SEDML}
\end{figure*}

We show in Fig.~\ref{fig:SEDML} (left panel) how the SED of a source with $L_{\rm {bol}} \sim $ 10$^{45}$ erg/s can be modified if the accretion disc is (de)-magnified due to microlensing by typically $\Delta m \sim $ 0.6\,mag at $\lambda \sim 1\,\micron$. We have picked two particular events, but the situation remains very similar if random events are selected. Although this amount of microlensing is relatively large, it is not uncommon and is indeed observed in several lensed systems \citep{Sluse2012b}. In this example, we have assumed an extended torus and our fiducial value of $f_{\rm{AD}}$. Because of the assumed source luminosity and extended torus model, only the accretion disc is significantly microlensed. We see that in case of demagnification, the peak of the the SED at $\sim 2.2\,\micron$ gets more pronounced because of the increased contrast between the torus and the disc, while in case of magnification, the contrast is decreased and the bump in the SED gets nearly suppressed. For an even larger magnification, this bump can disappear completely. The right panel of Fig.~\ref{fig:SEDML} shows the ratio between the microlensed and unlensed SED. This is qualitatively similar to the chromatic changes which would be observed in the flux ratio between 2 lensed images, one being much more affected by microlensing thant the other. It is ubiquitous that the main change of the SED takes place below $\lambda_{\rm {rest}}=2\,\micron$. The changes of the SED depend on $f_{\rm{AD}} (\lambda)$ but also to some extent on the characteristics of the microlensing event. This precludes the direct use of ratios of SEDs of different lensed images to infer $f_{\rm{AD}}$. The time variation of the event might help to characterise the microlensing but the monitoring time scales needed are discouragingly large (see Sect.~\ref{subsec:timescale}). 

The main interest of studying a microlensed rather than an unlensed quasar relies in the change of contrast between the torus and the disc emission induced by microlensing. The most useful targets in this context are quadruply imaged systems. Since microlensing is independent in each lensed image, it is likely, that at a given time, one of the lensed image will be affected by large microlensing while another one will only be weakly microlensed. The estimate of the amplitude of microlensing of the accretion disc in the images can be based on spectroscopy, comparing the flux ratios in the continuum and in the broad emission lines of the different pairs of lensed images \citep{Sluse2012b, Sluse2011a, Hutsemekers2010}. Alternatively, a comparison of the SED of the different lensed images could also be informative since similar SEDs suggest low microlensing (de)magnification. The most microlensed image, in case of significant demagnification of the accretion disc emission, could be used to study the emission from the torus. In that case, microlensing acts as a natural coronograph, hiding a fraction of the disc emission. On the other hand, in the case of significant magnification of the accretion disc emission, study of the intrinsic shape of the disc will be possible up to longer wavelengths. Finally in a situation of monitoring of a micro-magnification event, comparison of the observed signal with microlensing simulations could be used to derive the intrinsic disc temperature profile up to rest-NIR wavelengths \citep[e.g.][]{Anguita2008a, Eigenbrod2008b, Blackburne2011a}.

\section{Conclusions}
\label{sec:conclusions}

We have studied microlensing of large sources taking place in images of strongly-lensed quasars when observed in the near-/mid-infrared wavelength range (typically from 2 to 11\,$\micron$). Our simulations consider objects with intrinsic bolometric luminosities{\footnote{It is important to bear in mind that the accretion disc source size depends also on the black hole mass. Because we have assumed a constant Eddington ratio $L/L_{\rm{edd}} = 1/3$, we could express our results as a function of the source luminosity only. Therefore, our results indirectly depend on the Eddington ratio. }} in the range $10^{44.2}-10^{46}$\,erg/s. Our two-component model of the quasar emission allows us to distinguish between microlensing of the dust torus, whose size scales as $L^{0.5}$, and of the accretion disc, whose size scales as $L^{2/3}$ (for a fixed Eddington ratio). We find that microlensing with amplitude $\geq$ 0.1\,mag is expected to be common in the considered wavelength range. Most of the expected signal is due to microlensing of the accretion disc. The latter cannot be neglected as soon as the disc contributes to more than 5\% of the total flux, a situation expected to be common. On the other hand, microlensing of the torus becomes noticeable for the lowest luminosity quasars we considered, and at rest-wavelengths typically $\leq 2.2\,\micron$, i.e. where the torus is still compact and emission is dominated by the hot ($T \sim 1400$\,K) dust component. We may expect targets with $L_{\rm {bol}} \lesssim 10^{44}$\,erg/s to show even stronger microlensing of the torus, offering the perspective of studying the latter with microlensing. Such faint systems are however uncommon among the known lensed AGNs. This might change with upcoming all sky surveys like LSST or through dedicated lensed-quasars hunting programmes{\footnote{Any such programme, if based on colour selection criteria, will have to account for colour changes induced by microlensing to be efficient.}}.

It is well known that the probability distribution of microlensing of a compact source is different for images which are saddle-points or minima of the arrival-time surface \citep{Schechter2002, Saha2011a}, the former being more prone to demagnification than the latter. From the analysis of microlensing of the torus, we could see that the differences are less significant for more extended sources. More interestingly, the magnification distribution of the compound source is very different from the one obtained for the disc alone. The main characteristic of the PDF of the compound model is the suppression of large demagnifications. The reason of this suppression is simply that, when the disc is demagnified, its flux gets diluted in the total flux to such extent that only a weak demagnification is observed. If $f_{\rm{AD}}$ is the fraction of flux from the disc, and if the torus is assumed to be unlensed, the maximum demagnification of the compound model is given by $\Delta m = 2.5\,\log(1-f_{\rm{AD}})$. Conversely, in case of magnification, the contrast between the torus and the disc decreases, and large amplitude microlensing remains possible.

One of our aims was to estimate if microlensing could explain flux ratio anomalies in the NIR-MIR. For that purpose, we have calculated the expected distribution of $R_{\rm {cusp}}$ for quasars of different luminosities, assuming a cusp system similar to J1131$-$1231 \citep{Sluse2003}. We found that for almost all the situations considered, $R_{\rm {cusp}}$ may get significantly perturbed due to microlensing. A possibility to identify if an anomaly is caused by substructure or not might be through the study of the spectral energy distribution (SED) of the lensed images up to 4$\,\micron$ (rest-frame). We have shown that in case of important microlensing, the SED could be  significantly modified. This suggests that the comparison of the SED of the different lensed images should allow one to disentangle between microlensing and milli-lensing. In case of milli-lensing by substructures of mass of at least $10^5 M_{\sun}$, the dust torus should be significantly magnified (because the Einstein radius of the substructure is  similar to the size of the near-infrared torus), leading to less severe deformations of the SED. An estimate of the differential effects between the accretion disc and the torus (and within the torus) in the case of milli-lensing is needed to assess the absence of change of the SED in that case. Another possibility to disentangle microlensing from milli-lensing is to measure flux ratios at rest-frame wavelengths larger than typically 10 $\micron$, where the emission is dominated by the extended cold dust component, and the fraction of flux originating from the disc drops in general below 5\%. Measuring flux ratios in that wavelength range will be possible with the James Webb Space Telescope (JWST).

\begin{acknowledgements}
We thank the referee for his/her suggestions which improved the quality of the paper, and Prof. C.S. Kochanek for useful comments. DS acknowledges support from the Deutsche Forschungsgemeinschaft, reference SL172/1-1. OW is partly supported by the Emmy-Noether-Programme of the Deutsche Forschungsgemeinschaft, reference WU 588/1-1.
\end{acknowledgements}

\bibliographystyle{aa}
\bibliography{/Users/sluse/work/articles/bibds}

\end{document}